\documentclass[twocolumn,prb,aps,amsmath,amssymb,floatfix,showpacs,longbibliography,superscriptaddress]{revtex4-1}

\usepackage{color}
\usepackage{graphicx}
\usepackage{dcolumn}
\usepackage{bm}
\usepackage{subfigure}
\usepackage{array}
\newcolumntype{L}{>{\centering\arraybackslash}m{0.33\textwidth}}
\usepackage[colorlinks=true,linkcolor=blue,citecolor=blue,urlcolor=blue]{hyperref}

\graphicspath{{../figs/}}
\begin{document}

\newcommand{\bo}{\boldsymbol}
\newcommand{\boq}{\mathbf{q}}
\newcommand{\bok}{\mathbf{k}}
\newcommand{\bor}{\mathbf{r}}
\newcommand{\boG}{\mathbf{G}}
\newcommand{\boR}{\mathbf{R}}
\newcommand\2{$_2$}

\newcommand\myaff{nanomat/QMAT/CESAM and European Theoretical Spectroscopy Facility
Universite de Liege, Allee du 6 Aout 19 (B5a), 4000 Liege, Belgium}

\title{Remote free-carrier screening to boost the mobility of Fr\"ohlich-limited 2D semiconductors}

\author{Thibault Sohier}
\affiliation{\myaff}
\author{Marco Gibertini}
\affiliation{Dipartimento di Fisica Informatica e Matematica, Universit\`a di
Modena e Reggio Emilia, Via Campi 213/a, I-41125 Modena, Italy}
\author{Matthieu J. Verstraete}
\affiliation{\myaff}

\date{\today}


\begin{abstract}
Van der Waals heterostructures provide a versatile tool to not only protect or control, but also enhance the properties of a 2D material.
We use ab initio calculations and semi-analytical models to find strategies which boost the mobility of a current-carrying 2D semiconductor within an heterostructure.
Free-carrier screening from a metallic ``screener'' layer remotely suppresses electron-phonon interactions in the current-carrying layer.
This concept is most effective in 2D semiconductors whose scattering is dominated by \textit{screenable} electron-phonon interactions, and in particular the Fr\"ohlich coupling to polar-optical phonons.
Such materials are common and characterised by overall low mobilities in the small doping limit, and much  higher ones  when the 2D material is doped enough for electron-phonon interactions to be screened by its own free carriers.
We use GaSe as a prototype and place it in a heterostructure with doped graphene as the ``screener'' layer and BN as a separator.
We develop an approach to determine the electrostatic response of any heterostructure by combining the responses of the individual layers computed within density-functional perturbation theory.
Remote screening from graphene can suppress the long-wavelength Fr\"ohlich interaction, leading to a consistently high mobility around $500$ to $600$ cm$^2$/Vs for carrier densities in GaSe from $10^{11}$ to $10^{13}$ cm$^{-2}$.
Notably, the low-doping mobility is enhanced by a factor $2.5$.
This remote free-carrier screening is more efficient than more conventional manipulation of the dielectric environment, and it is most effective when the separator (BN) is thin.
\end{abstract}

\maketitle

\section{Introduction}

Van der Waals heterostructures (VdWh) are becoming a device design paradigm in 2D materials applications\cite{Li2017a,Liang2020}.
The operating layer, performing the primary functionality, is included in a stack of other 2D layers fulfilling secondary roles like protection, gating or control.
Encapsulating 2D materials in boron nitride (BN), for example, has already proven to be highly beneficial to the quality and cleanliness of the operating material's response\cite{Dean2010,Cadiz2017}.
The exciting prospect of including supporting 2D layers, to engineer the properties of the operating material beyond its intrinsic limits, has been much discussed in the past decade\cite{Geim2013,Novoselov2016a} but is only starting to be realized\cite{Tielrooij2018,Rivera2018,Wakamura2018,Kim2020}. 
With this aim, one must understand, control, and exploit the interactions between all the layers within a VdWh.
The present work takes a critical step towards this challenging task, in taking fully and quantitatively into account the mutual dielectric feedback between 2D layers.

VdWh engineering brings particularly interesting opportunities to electronic transport.
High-mobility semiconductors are useful in many devices, especially when coupled with electrostatic doping\cite{Perera2013,Riederer2018,Zhao2019}, which allows to explore a wide range of carrier densities, in a non-destructive and versatile way. In this context, depending on the application and the situation, the operating layer needs to perform well in many different doping regimes (hereafter, the nature of the doping should be understood as electrostatic).
As discussed in the literature \cite{Ma2014a,Sohier2020}, it is a strong challenge for materials to display consistently good mobilities over a large range of carrier densities.
This is particularly true for 2D materials whose scattering is dominated by the Fr\"ohlich interaction with polar-optical phonons.
This work explores the possibility of exploiting their integration in a VdWh to provide uniform performance over a range of doping levels.

Ab initio simulation of transport properties has shown promise in its ability to guide materials design.
However, performing such studies for materials within a VdWh and over a large range of carrier densities remains   a challenge.
VdWh are difficult to simulate ab initio due to their multiple periodicities, entailing simulation supercells which are prohibitively large.
The simulation of doping is also not obvious:
Most ab initio electron-phonon scattering calculations in semiconductors are done in the zero doping limit.
The ability to self-consistently simulate electron-phonon interactions in electrostatically doped 2D materials was recently developed\cite{Sohier2017}, but remains computationally affordable only when Fermi surfaces are large enough, i.e.\, at large enough doping.
Models to bridge the gap between the zero and large doping regimes are still lacking.

Here we propose a step towards ab initio simulations of transport in VdWh devices over a large range of doping.
In particular, we propose a framework to deal with ``screenable'' contributions to electron scattering by phonons, as those are likely to be most affected by VdWh integration and doping.
We further focus on one of the most common and important type of screenable electron-phonon coupling: the Fr\"ohlich interactions with polar-optical phonons.
A semi-analytical scheme is used to treat the electrostatics of the VdWh including dielectric and free-carrier screening from different layers.
The response of each individual layer to a generic potential perturbation is computed in Density Functional (Perturbation) Theory (DFT / DFPT), then they are combined in a model for the full response of the VdWh to the Fr\"ohlich potential(s).
Dielectric models of VdWh have been developed in the past \cite{Andersen2015,Thygesen2017,Gjerding2020,Guo2020}, with a focus on quantities like the dielectric function and exciton binding energies.
We propose another formal framework, with a focus on the potentials generated by electron-phonon interactions, and apply it to a new problem: phonon-limited electronic transport.

This VdWh electrostatics model is used to demonstrate a solution to the aforementioned materials design challenge, i.e.\ high mobility over a wide range of doping regimes, using the prototypical example of GaSe.
In a recent work \cite{Sohier2020}, the outstanding transport performance of GaSe was predicted in the high doping regime.
Transport in this material is limited by the Fr\"ohlich interaction, and the high mobility at high doping relies largely on the screening of this interaction by free carriers added to GaSe.
This is confirmed in the present work by showing a decrease of intrinsic mobility by more than a factor $3$ in the low doping regimes.
It is then shown that this decrease can be avoided by putting GaSe in proximity with a metal, providing free-carrier screening externally, irrespective of GaSe's doping.
In particular, we propose a GaSe/BN/graphene heterostructure, with doped graphene as a ``screener'' layer and BN as a separator.
The principle of proximity free-carrier screening has recently been used to tune the band gap of semiconductors\cite{Qiu2019} and to manipulate electron-electron interactions in graphene\cite{Kim2020}.
We use it here to engineer electron-phonon interactions within the current-carrying 2D semiconductor.

This paper is structured as follows.
In section \ref{sec:Frolim}, we use the case of GaSe to discuss the doping-dependent performance of 2D layers in which the Fr\"ohlich interaction dominates electron-phonon scattering, and demonstrate that the lack of intrinsic free-carrier screening at low doping leads to very low mobility.
In section \ref{sec:model}, we develop the electrostatic model that allows us to calculate the response of the full VdWh from the ab initio response of each layer computed independently.
Finally, in section \ref{sec:results} we apply this approach to the GaSe/BN/Graphene heterostructure and show that the mobility can be kept at a high value between $500$ and $600$ cm$^2$/Vs for carrier densities from $10^{11}$ to $10^{13}$ cm$^{-2}$.

\section{Fr\"ohlich limited 2D semiconductors}
\label{sec:Frolim}

\begin{figure}[h]
   \includegraphics[width=0.47\textwidth]{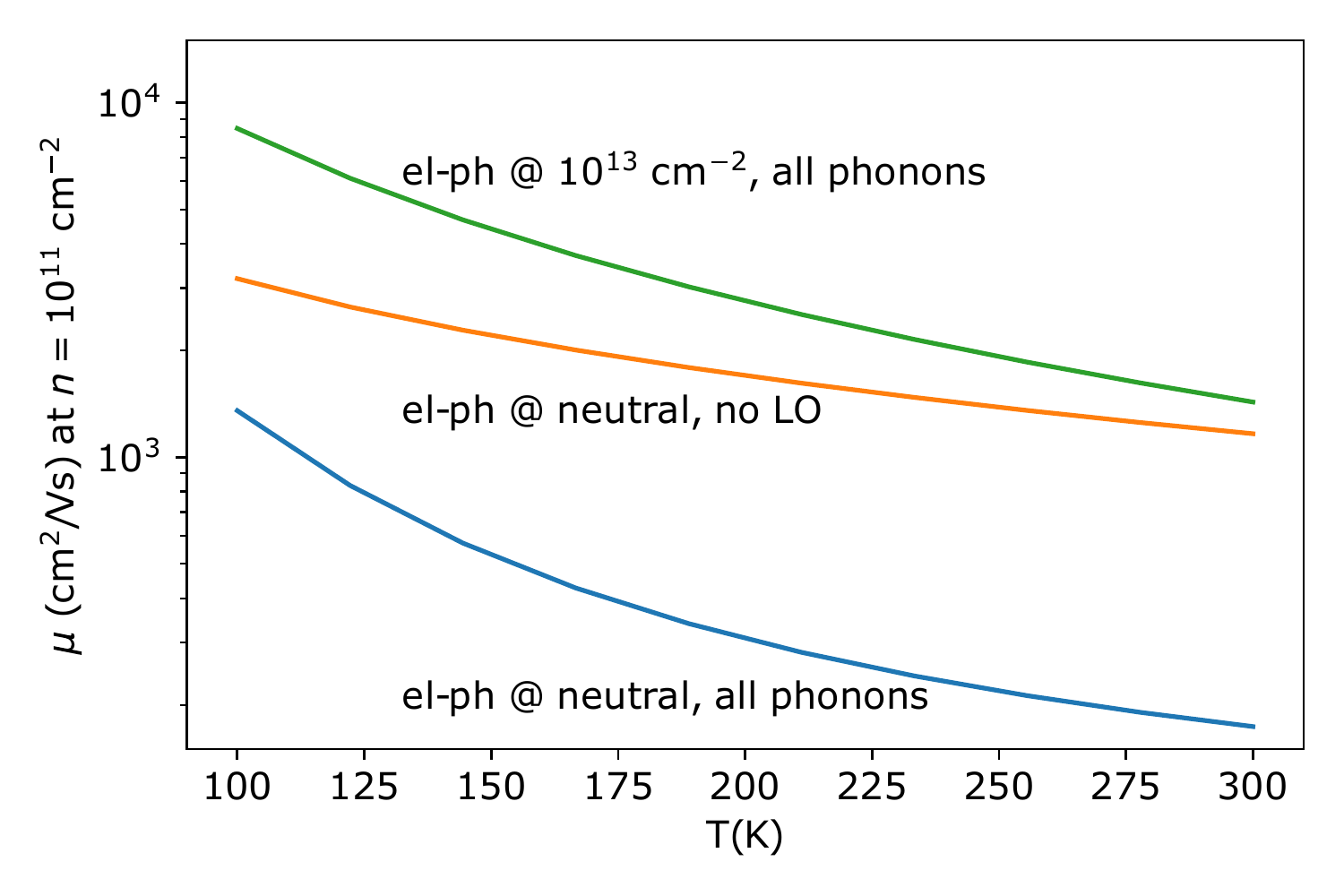}
   \caption{Ab initio mobility versus temperature in neutral GaSe, in the $n \to 0$ limit (reached at $n \simeq 10^{11}$).
   The mobility without the longitudinal optical (LO) phonons is plotted to show that scattering with this mode limits transport.
   Finally, we also compute mobility in a fictitious system by using the EPIs of GaSe doped at $n = 10^{13}$ cm$^{-2}$ while simulating the transport at $n = 10^{11}$ cm$^{-2}$.
   This shows that free carrier screening of the EPIs, induced by doping the layer itself, is able to suppress the main electron-phonon scattering mechanisms and increase the mobility.}
   \label{fig:mob_neut}
\end{figure}

\begin{figure}[h]
\includegraphics[width=0.47\textwidth]{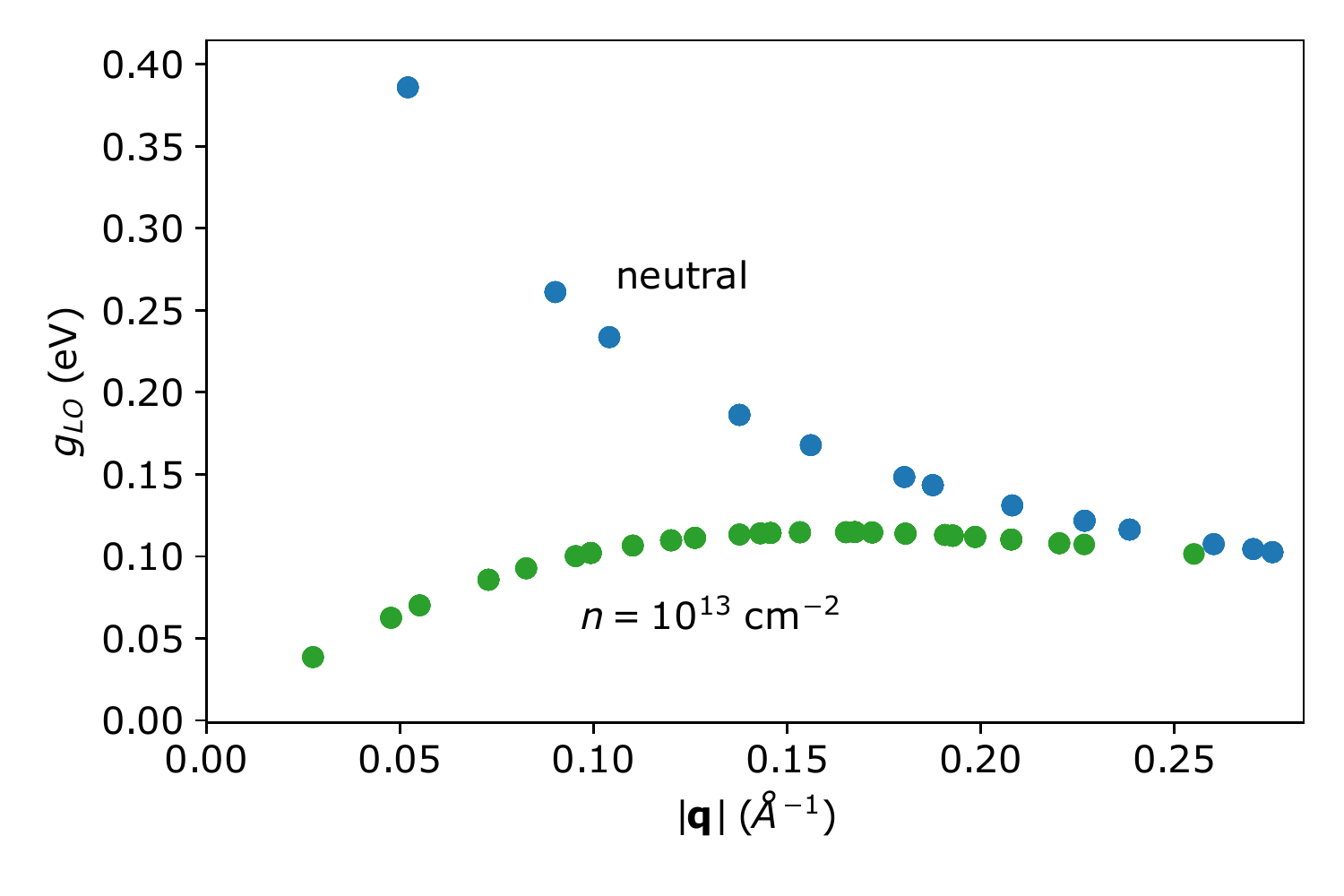}
\caption{EPIs between the LO mode and states at the bottom of GaSe's conduction band, computed within DFPT for neutral and doped GaSe, as a function of the phonon momentum.
Other modes have non-negligible coupling (LA and A$_{\rm 1g}$), but LO clearly dominates scattering through the Fr\"ohlich interaction.
It is strongly affected by free-carrier screening: in the doped case, the coupling vanishes as $\Gamma$ and overall it barely reaches a $10\%$ of the maximum value in the neutral case ($\sim 1.124$ eV at $\Gamma$).}
  \label{fig:g_neut_dop}
\end{figure}

The broad class of materials concerned by this work are 2D semiconductors for which the electronic transport performance is limited by scattering mechanisms which are sensitive to free-carrier screening.
It is not obvious to further qualify this class as a whole, which hosts a variety of different members.
Focusing on intrinsic scattering mechanisms driven by electron-phonon interactions (EPIs), multi-valley materials can usually be excluded, since inter-valley EPIs are often strong\cite{Sohier2018,Sohier2019}, and at momenta larger than the size of the Fermi pocket that characterizes the free carriers providing the screening.
In this large momentum regime, free-carrier screening is inefficient even if the EPI are sensitive to it.
In single valley materials (up to quite high chemical potentials), it is reasonable to assume that free-carrier screening will be efficient on screenable EPIs.
These include the Fr\"ohlich, piezoelectric, and acoustic ``deformation-potential'' EPIs. Others, like those responsible for graphene's intrinsic transport properties, are altogether insensitive to screening. Which kind of EPI dominates transport will depend on the specific material.
This work focuses on 2D materials in which the dominant EPI is the Fr\"ohlich interaction between electrons and polar-optical phonons, usually the longitudinal optical (LO) modes.
Fr\"ohlich EPIs concerns any semiconductor in which the atoms of the unit-cell carry different Born effective charges (BECs).
This includes any non-elemental material, but also elemental materials with some factor disrupting the balance between the atoms \cite{Bistoni2019}.
Fr\"ohich EPIs increase as: BECs increases, screening decreases, or the LO phonon energy decreases.
Since they are mediated by long-range electric fields, they are screenable.
It is certainly one of the most pervasive and critical source of EPI\cite{Ponce2020,Ma2020}, and it has been extensively modelled in both 3D \cite{Mori1989, Sarma1985, Vogl1976, Sjakste2015, Verdi2015} and 2D \cite{Sohier2016}, with parameters for the BECs and the dielectric properties.
In 3D bulk materials, the Fr\"ohlich EPI diverges as the inverse of the phonon momentum in the long wavelength limit \cite{Ponce2015}, which can have a strong impact on scattering.
While the EPI stays finite in 2D systems, it still undergoes a sharp increase at small momenta, and the Fr\"ohlich EPI can easily dominate all other mechanisms as in 3D.

In the 2D framework, the long wavelength Fr\"ohlich electron-phonon coupling (see App.\ \ref{app:Frolich}) can be thought of as the ratio of a parameter depending on BECs and mildly on momentum, and the dielectric function $\epsilon(q)$, which accounts for both the environment and the material containing the electrons involved.
The dielectric function  in the long wavelength limit ($q\to0$) can be modeled as $1+ \alpha q$ for a 2D material in vacuum,  where $\alpha$ is the polarizability of the 2D layer~\cite{Cudazzo2011,Latini2015,Trolle2017,Tian2020}, but the present work relies on a more detailed and realistic model.
One general behavior is that, in 2D, the dielectric function is dominated by the response of the environment for $q \to 0$ and by that of the 2D material for $q \to \infty$.

In a recent work \cite{Sohier2020}, eleven of the best conductors within a database of exfoliable materials\cite{MC2D,Talirz2020,Mounet2018} were identified.
Seven of them (GaSe,  InSe, Bi$_2$SeTe$_2$, Sb$_2$SeTe$_2$, BiClTe, AlLiTe$_2$, BiSe$_3$) display rather large BECs and strong Fr\"ohlich EPIs. Since the calculations were done at a relatively large doping, these EPIs were screened by local free carriers, and did not affect drastically the conductivity.
However, those same materials can be expected to have much lower mobilities at low doping, when screening is ineffective.

A prototypical example is GaSe (see Ref.~\onlinecite{MC2D} for basic properties).
For reference, the room temperature mobility at $n=10^{13}$ cm$^{-2}$ was computed to be $\mu \simeq 600$ cm$^2$/Vs \cite{Sohier2020}, placing it among the very best performing 2D semiconductors at high doping.
We now look at electronic transport in GaSe in the zero doping limit.
EPIs are computed in DFPT and the full energy- and momentum-dependent Boltzmann transport equation is solved iteratively as described in Ref.~\onlinecite{Sohier2018}.
To compute the zero doping limit, we use $n=10^{11}$ cm$^{-2}$, as we found that the mobility is stable within $~ 2\%$ below that value.
The mobility as a function of temperature is shown in Fig.\ \ref{fig:mob_neut} for 3 different sets of EPIs (2 of them fictitious).
In the first (realistic) system, we use the EPI matrix elements as computed in the neutral material ($n=0$).
This represents the standard small doping limit.
The room-temperature mobility is $\mu \simeq 174$ cm$^2$/Vs, much lower than the high doping value.
For the second (fictitious) system we use the same EPI, but without the Fr\"ohlich-inducing phonon LO.
The mobility increases by an order of magnitude, clearly showing that Fr\"ohlich is limiting the mobility.
In the third system, we use the EPI computed in GaSe at a doping of $n=10^{13}$ cm$^{-2}$ in Refs. \onlinecite{Sohier2020,MCArchive}.
This system is fictitious because the BTE is solved with a doping level ($n=10^{11}$ cm$^{-2}$) different from the one used in EPI calculations ($n=10^{13}$ cm$^{-2}$).
It is instructive since it shows that, keeping all other factors the same, using the EPIs from the doped system leads to an order of magnitude increase in the mobility.
This is due to the screening  of the Fr\"ohlich EPI  by the high density of added free carriers, as confirmed in Fig.\ \ref{fig:g_neut_dop} showing the coupling of the LO mode $g_{LO}(|\boq|)$ for the neutral and doped systems.
It can thus be inferred that free-carrier screening of the Fr\"ohlich EPI would enhance the mobility of GaSe in the low doping limit, similarly to how screening affects carrier relaxation in a related material, InSe~\cite{Chen2020}.
Since there are not enough intrinsic free carriers in GaSe in the low doping regime, one would need an external source of free-carrier screening.
We propose to place GaSe in a VdWh with doped graphene to screen the Fr\"ohlich EPIs remotely.
To demonstrate this, we first develop a model for the complete electrostatics of such systems.

\section{Van der Waals electrostatics model}
\label{sec:model}

This section describes a semi-analytical model parametrized with density-functional perturbation theory (DFPT) to solve the electrostatics (i.e.\ the response to a static electric field perturbation) of a VdWh in the presence of both dielectric and free-carrier screening.
We are especially interested in the response to the Fr\"ohlich potential generated by polar optical phonons.
The entire VdWh would be prohibitively expensive to simulate in DFPT, especially with the very fine wave vector grids needed for transport.
This model aims at re-constructing the full response of a VdWh from the response of each individual layer, which can be reasonably computed in DFPT.
The process relies on a response function model, which is isotropic and $q$ dependent in the plane and has a flexible profile in the out-of-plane direction.
Ultimately, we are interested in the screened potential felt by electrons in the operating layer, which will dictate its transport properties and the device performance.
More formal details can be found in App. \ref{app:model}. This model goes beyond previous analytical models meant to capture the first-order, long wavelength behavior of screening~\cite{Cudazzo2011,Latini2015,Trolle2017,Tian2020}, which are routinely employed~\cite{Keldysh1978,Cudazzo2011,Berkelbach2013,Sohier2017a}. Appendix \ref{app:AnaCompa} provides a more detailed comparison with such analytical models.

\begin{figure}[h]
\includegraphics[width=0.48\textwidth]{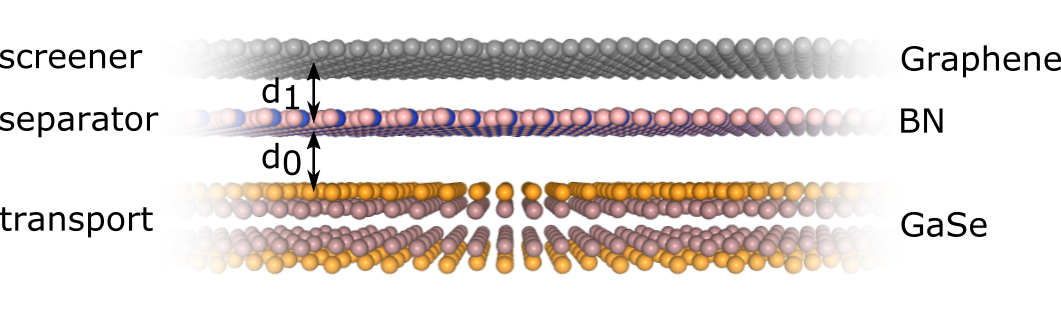}
\caption{Schematic view of the system studied. The transport layer (here GaSe or more precisely Ga$_2$Se$_2$, which is composed of two sublayers) is the operating 2D material that carries the current.
The screener layer (doped graphene) provides free-carrier screening remotely.
Finally, the separator (BN) electrically insulates the operating material and the screener.
The interlayer distances $d_0$ and $d_1$ can be different in general, but they are both fixed to $d=3.4$ \AA\ in this work.
Although monolayer BN is represented, multilayer BN is also considered.}
\label{fig:system}
\end{figure}

Although any combination of 2D layers can be studied, we focus here on GaSe/BN/Graphene systems, as represented in Fig.\ \ref{fig:system}.
Monolayer GaSe (gap $\approx 1.8$ eV  within the GGA-PBE approximation to the DFT exchange-correlation functional) is the operating material  in which electronic transport occurs.
Carrier densities ranging from the $n \to 0$ limit up to $n=10^{13}$ cm$^{-2}$ are considered in the GaSe layer.
Doped monolayer graphene is the ``remote screener'', with a fixed carrier density of $n=5\ 10^{13}$ cm$^{-2}$, which will be referred to later on via the notation Gr(doped).
BN has a large gap of $\approx 4.7$ eV in GGA-PBE and is routinely used in 2D VdWh as an encapsulator or gate dielectric.
It is necessary here to electrically isolate the operating material from the remote screener and avoid charge transfers.
Both monolayer and multilayer BN are studied in the following.

The electrostatics of the system are determined from the response of each individual layer to electric fields perturbations.
Given the in-plane periodicity and symmetry of the system, the corresponding perturbing potential is periodic in the plane, and written $V(q,z)$ where $q$ is the norm of the in-plane momentum, and $z$ the out-of-plane real-space variable.
Doped graphene responds like a metal, with perfect screening of in-plane electric fields in the long wavelength limit.
BN also brings a $q$-dependent dielectric response (inefficient at small $q$, and similar to bulk BN at large $q$).
The operating layer, GaSe, responds like a dielectric at low doping, and a metal at high doping.
The dielectric response of neutral GaSe is computed directly in DFPT, and the contribution of potential free-carriers (as induced in GaSe by electrostatic doping) is added on top.
As detailed in App.\ \ref{app:model}, the central quantity characterizing each layer's behavior is the interacting response function, assumed to be of the form:
\begin{align}\label{eq:chi}
\chi(q,z,z') =& Q(q) f(q,z-z_0)f(q,z'-z_0) \\
              &+ P(q) g(q,z-z_0)g(q,z'-z_0) \nonumber
\end{align}
$Q,f$ characterize the monopole contribution to the response, while $P,g$ represent the dipole part.
$f$ is an even function of $z$, and describes the normalized spatial profile of the material's response to a constant potential $V(q,z) \propto 1$.
$Q$ is the associated $q$-dependent amplitude of this response. Similarly, $g$ is odd and, along with the amplitude $P$, they represent the response to a linear potential $V(q,z) \propto z$ ($q$-dependent vertical  electric field).
Those functions are computed for each individual layer in DFPT for the range of momenta $q$ that eventually enter the Boltzmann transport equation.
Isotropy is assumed, so only one arbitrary direction is used for the momentum.

The responses of each layer are then combined in an electrostatic model of the VdWh, as detailed in appendix \ref{app:combi}.
This is done within the random-phase approximation (RPA)\cite{RPA1,RPA2,RPA3} in the limit of a negligible interlayer overlap~\cite{Guo2020}, so that each layer responds to an effective external potential made of the global external potential plus the sum of the induced potentials from all other layers.
This assumption allows us to define a simple system of equations that we then solve numerically.
In appendix \ref{app:tests}, the model is compared with direct DFPT calculations in mono- and bilayer BN to demonstrate its accuracy and the reliability of the assumption above.

Since they have finite Born effective charges, both BN and GaSe will generate Fr\"ohlich EPIs.
Electrons in GaSe thus couple to polar-optical phonons in both GaSe and BN, the latter being remote~\cite{Wang1972,Hess1979,Moore1980}.
Phonons are not explicitly simulated in the VdWh electrostatics model.
They are assumed to be unchanged from the isolated layers to the VdWh (no interlayer hybridization of modes).
The potential they generate is recreated from DFPT calculations, and used to perturb the VdW electrostatics model.
One improvement over previous models \cite{Sohier2016} is to exploit the parametrization of the layers' dielectric response to model the profile of the polarization density that generates the Fr\"ohlich EPI. The same profile is used as a proxy to project the Fr\"ohlich potential of electronic wave functions in order to obtain the corresponding electron-phonon coupling. Details of this model, as well as a comparison with direct DFPT calculations, are given in App.\ \ref{app:Frolich}.
Only first order dipole potentials are considered, quadrupole contributions \cite{Brunin2020a,Brunin2020,Jhalani2020,Park2020} are neglected.
In principle, multilayer materials generate several polar-optical phonons with different phase shifts in the layers\cite{Sohier2017a,Michel2011}.
Here we focus on the mode with largest Fr\"ohlich EPI, in which all layers are in phase.
We make the adiabatic approximation, allowing us to treat phonons as a static perturbation ($\omega = 0$).

The Fr\"ohlich potential and the responses are assumed to be isotropic in the range of $\boq$ vectors considered.
This is valid for all layers here, as is well-known for graphene and BN \cite{Sohier2016,Sohier2017a,Sohier2015}, and also in GaSe as we can see in Fig.~\ref{fig:g_neut_dop}.
Indeed, the $\boq$ vectors calculated within DFPT sample the whole Brillouin zone, along all possible directions; the fact that the scatter plot gives a line demonstrates isotropy.

Interlayer distances are chosen to be $3.4$ \AA, understood as the geometric distance between the outermost atomic planes of successive 2D materials.
For BN and graphene, there is only one atomic plane.
For GaSe, there are 4, the outermost being $2.4$ \AA\ away from the center of the layer.

For clarity, we compare our formalism with the existing Quantum Electrostatic Heterostructure (QEH) model\cite{Andersen2015,Gjerding2020} in App.\ \ref{app:QEHcompa}.
They both achieve a similar general purpose: to compute the dielectric response of a Van der Waals heterostructure from the ab initio response of each individual layer.
The main physical ingredients (monopole and dipole responses) and approximations (RPA and negligible interlayer hybridization/overlap) are the same. The parametrization from DFPT and the scope of application differ. The difference in application means we perturb with different potentials and extract different quantities.
If we were to apply exactly the same perturbation (bare Fr\"ohlich potentials) and extract exactly the same quantities (screened Fr\"ohlich potentials), we expect the different parameterization to yield a qualitatively similar result. The extent of the quantitative numerical difference is difficult to estimate.

Ab initio calculations of structures, ground states and dielectric responses are performed with Quantum ESPRESSO\cite{Giannozzi2009,Giannozzi2017} (QE).
Full electron-phonon interactions and transport calculations in neutral and doped GaSe were done for comparison, with ultrasoft pseudopotentials from the SSSP library\cite{Prandini2018a} (efficiency version 0.7).
The Phonon code of QE has been modified to compute the dielectric response of each layer.
More specifically, the phonon perturbation is replaced by the potentials in Eq.~\eqref{eq:Vpert} of App.~\ref{app:resp}.
Those modifications are similar to a previous work \cite{Sohier2015}, with the addition of the dipole perturbation.
Dielectric responses were computed using optimized norm-conserving Vanderbilt pseudopotentials\cite{Hamann2013} from the pseudo-Dojo library\cite{VanSetten2018}, as the modifications of the Phonon code have not been implemented yet for other types of pseudopotentials.
We use the AiiDA materials informatics infrastructure\cite{Pizzi2016,Huber2020} to manage calculations and store data.
The solution of the VdWh electrostatics model is implemented in Python. The associated code, along with a database containing the electrostatic responses of doped graphene, BN and GaSe is available \footnote{\url{https://gitlab.com/tsohier/vdw_electrostatics}}.

\section{Results}
\label{sec:results}

\begin{figure}[h]
   \includegraphics[width=0.45\textwidth]{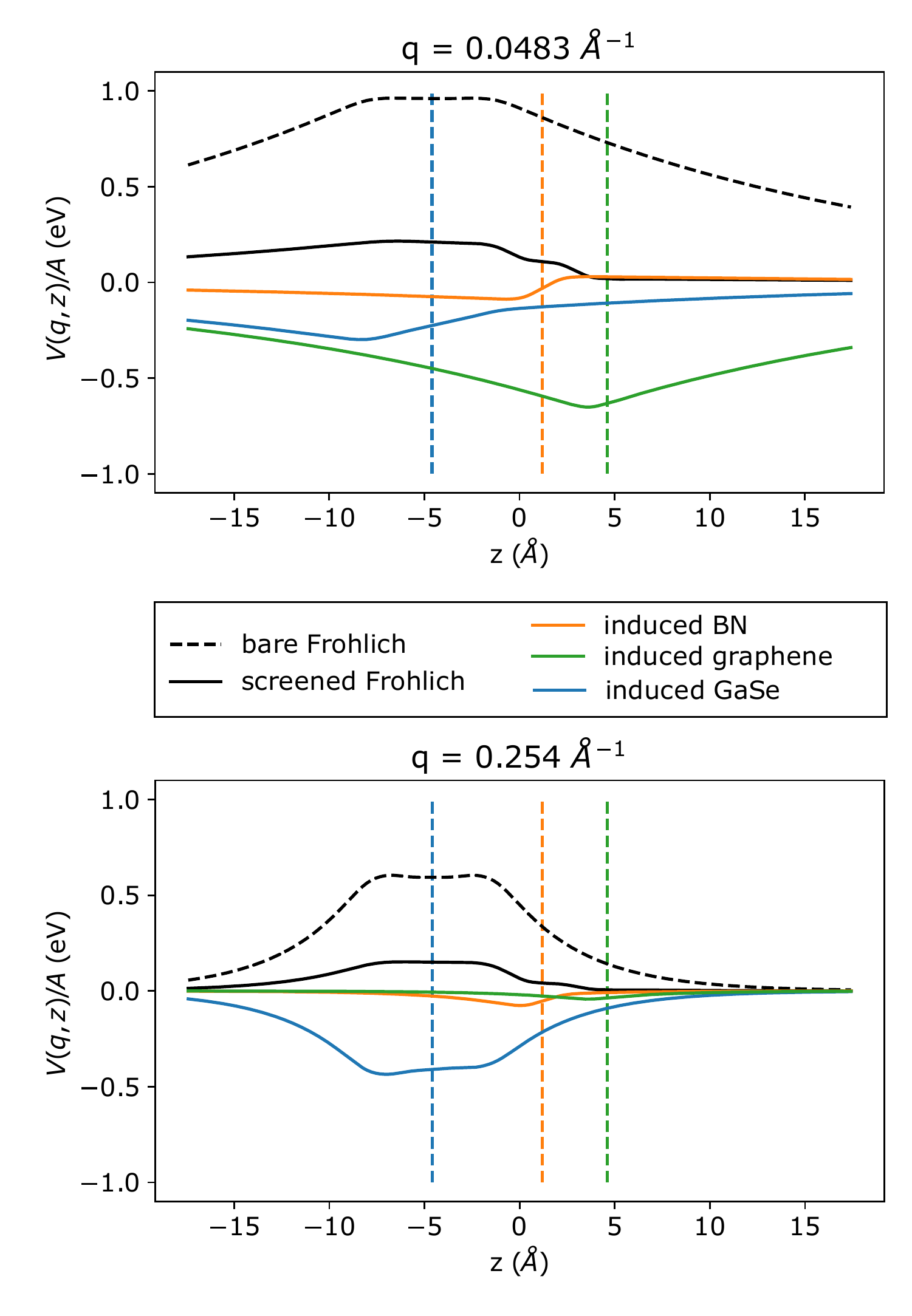}
   \caption{Bare, induced and screened potentials for the most relevant configuration: GaSe(neutral)/BN/Gr(doped), in response to the Fr\"ohlich potential generated by GaSe, at two values of momentum.
   The induced potential is separated  into contributions from each layer, with their position indicated by the vertical dashed lines. The definition and units of $V(q,z)/A$ correspond to those of the Fr\"ohlich potential given in appendix \ref{app:Frolich}.}
   \label{fig:pot_soup}
\end{figure}

Our approach provides a clear and intuitive physical understanding of the VdWh in terms of electrostatics.
Fig.~\ref{fig:pot_soup} illustrates this by showing the potentials solving the model in the most relevant configuration: GaSe(neutral)/BN/Gr(doped) perturbed by a Fr\"ohlich potential from GaSe's polar-optical phonons.
We select two momenta at the extrema of the interval considered in this work.
At small $q$, graphene's response is prominent.
Indeed, the perturbation extends far in the out-of-plane direction and graphene is metallic, so the induced potential is (negatively) large.
We also note that since the bare Fr\"ohlich potential varies over the dielectric thickness of the graphene layer, graphene's response is not symmetric with respect to the middle of the layer.
Responses from BN and GaSe both display clear dipole-like features with positive and negative electric field regions.
For GaSe, the dipole component originates mostly from the response to graphene's induced potential, which varies significantly over GaSe's thickness, yielding a finite electric field.
Despite the dipolar feature, GaSe's response clearly does not average to zero, indicating that there is a significant monopole component to its response as well, this time triggered mostly by the bare Fr\"ohlich perturbation.
BN's response also includes both monopole and dipole components, but is closer to a purely dipolar one.
BN feels two main potentials with finite derivatives: the bare Fr\"ohlich from GaSe and the induced response from graphene.
They have counteracting effects, and the sign of the dipolar response confirms that the bare Fr\"ohlich from GaSe dominates.

At large $q$, GaSe itself performs most of the screening, with a small contribution from BN on one side.
The induced potential from graphene is very weak compared to the smaller $q$ case, due to the fact that the bare Fr\"ohlich potential decays more rapidly (as $~ e^{-q|z-z_{\rm{GaSe}}|}$) in the out-of-plane direction and doesn't reach graphene.
The screened Fr\"ohlich potential is slightly smaller (more screened) on the side of GaSe adjacent to BN.
This was true for the small $q$ case as well, and is a general and intuitive feature: the screening is more efficient towards the separator and screener layers.
However, the difference is within $2\%$: despite the asymmetry of the induced potentials (with respect to their respective layers), the total screened potential ends up quite flat within each layer.
This indicates that the dipole response of each layer is near-perfect, meaning that any perpendicular electric field is almost fully compensated.
\\

\begin{figure}[h]
   \includegraphics[width=0.45\textwidth]{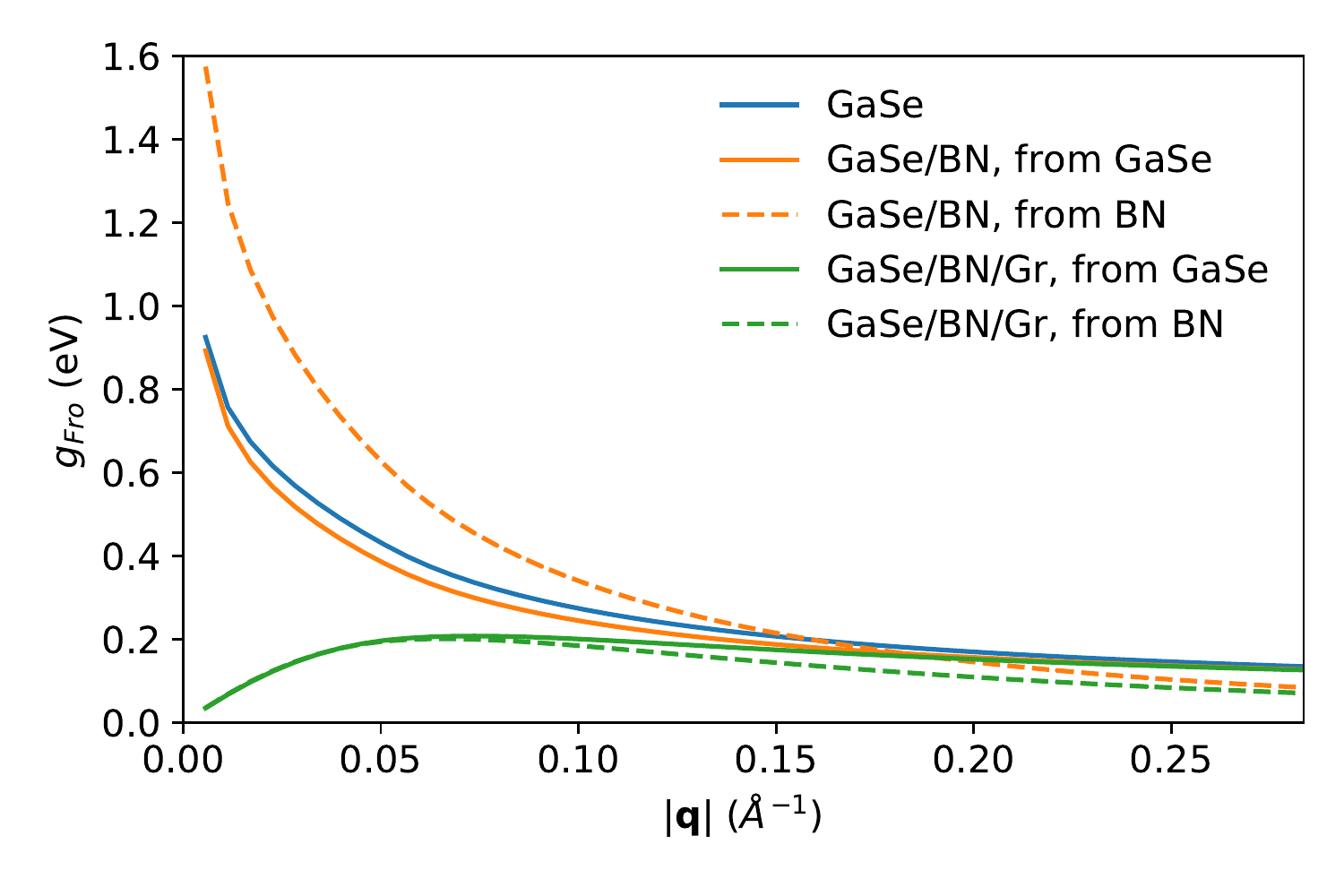}
   \caption{Fr\"ohlich EPIs from polar-optical phonons in both GaSe (plain) and BN (dashed), as felt by electrons in GaSe, in different setups: GaSe(neutral) alone, GaSe(neutral)/BN, and GaSe(neutral)/BN/Gr(doped), with at $n=5\ 10^{13}$ cm$^{-2}$ in graphene.
   Fr\"ohlich potentials coming from both GaSe and BN are considered, but always ``as felt by GaSe electrons'', that is, the Fr\"ohlich potential is always averaged over the GaSe layer.}
   \label{fig:remote_screening}
\end{figure}

In Fig.\ \ref{fig:remote_screening}, the electrostatics are solved in different systems, gradually adding the key layers and observing the effects on the screened Fr\"ohlich EPIs as felt by GaSe's electrons. Those are computed by averaging the full screened potential over the GaSe layer, as detailed in App.\ \ref{app:Frolich}.
First, in neutral GaSe alone, the Fr\"ohlich potential is only screened dielectrically by GaSe.
Second, in GaSe(neutral)/BN, there is a relatively weak additional dielectric screening from BN: this represents the standard type of screening one can expect from a dielectric environment (substrate or encapsulator).
An additional remote Fr\"ohlich potential~\cite{Wang1972,Hess1979,Moore1980} from BN comes into play.
Although it is quite strong, its consequence on transport is limited by the fact that the energy of the associated phonon is very large ($\sim 0.19$ eV).
Indeed, injecting this coupling into a simple Fermi golden rule, the scattering rate of a state at the Fermi level is proportional to
$n_{\rm{BE}}(\hbar \omega) \times (1-n_{\rm{FD}}(\varepsilon_F+\hbar \omega))+(n_{\rm{BE}}(\hbar \omega)+1)\times(1-n_{\rm{FD}}(\varepsilon_F-\hbar \omega))$
where $n_{\rm{BE}}$ and $n_{\rm{FD}}$ are the Bose-Einstein and Fermi-Dirac occupation functions for phonons and electrons, respectively.
Up to high temperature, there will be few phonons to absorb ($n_{\rm{BE}}(\hbar \omega) \ll 1 $), and phonon emission (second term) will be limited by the fact that states at $ \varepsilon_F - \hbar \omega$ are mostly occupied ($1-n_{\rm{FD}}(\varepsilon_F-\hbar \omega) \ll 1$).
In fact, based on LO frequencies in GaSe and BN ($200$ and $1500$ cm$^{-1}$, respectively), this expression allows to estimate that for a equal EPI the scattering from BN's phonons will 3 orders of magnitude less efficient than GaSe's at room temperature.
Third, in GaSe(neutral)/BN/Gr(doped), the doped graphene sheet acts as a remote `` screener'' layer.
There is now some metallic screening, with the coupling vanishing as $q \to 0$.
The efficiency of this remote screening is limited in $q$,  as seen when comparing the induced potentials from graphene at small and large $q$ in Fig.\ \ref{fig:pot_soup}. The efficiency is related to the inverse of the distance between  graphene and GaSe.
Still, the remote screening is most essential in the critical low doping limit of GaSe, when the Fermi surface and $\mathbf{q}$ vectors relevant for transport in GaSe are small.
Thus, reducing the sharp increase at $q=0$ is enough to suppress the electron-phonon scattering.\\

\begin{figure}[h]
   \includegraphics[width=0.45\textwidth]{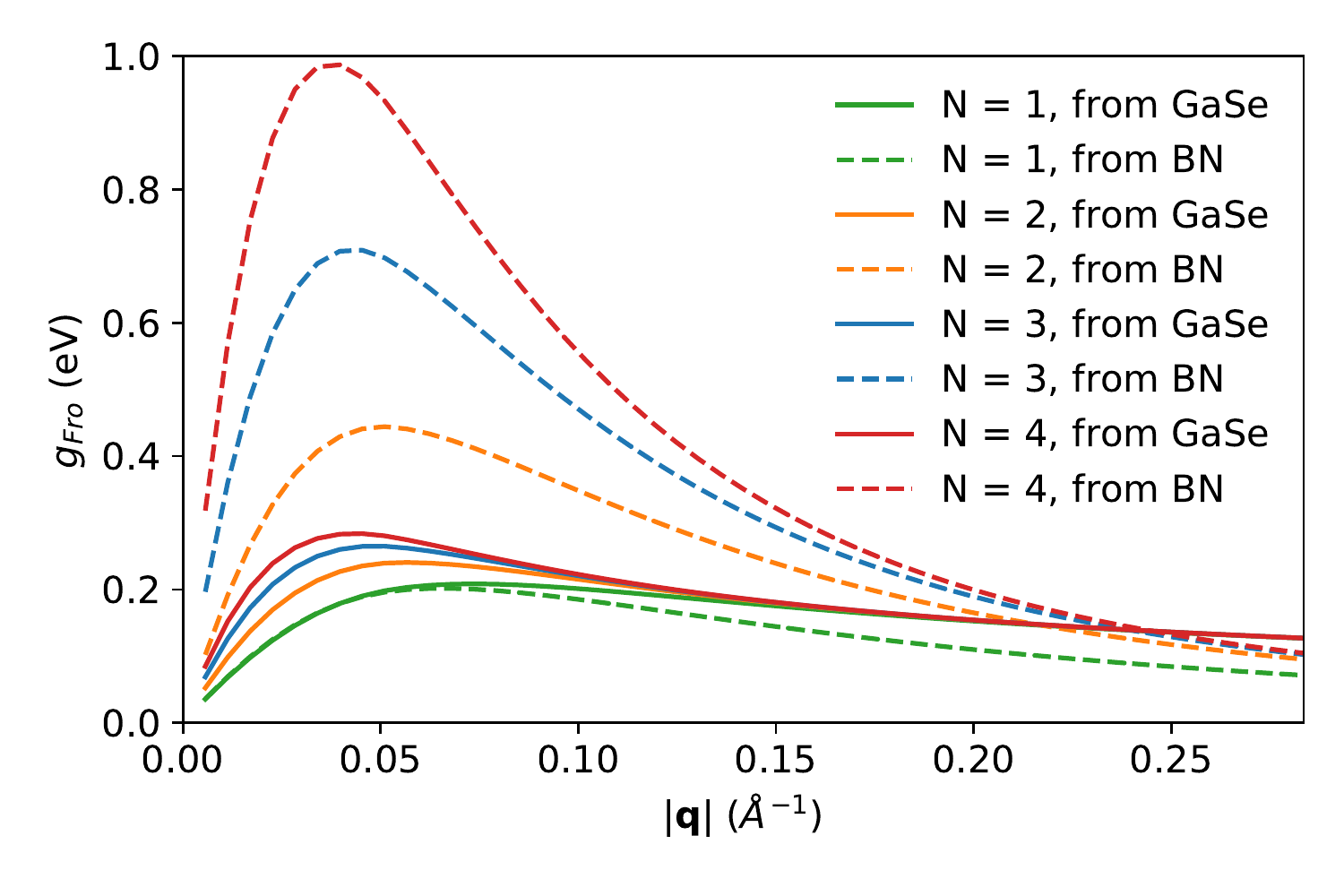}
   \caption{Fr\"ohlich EPI from polar-optical phonons in both GaSe (plain) and BN (dashed), as felt by electrons in GaSe, in GaSe(neutral)/BN/Gr(doped), changing the number of BN layers ($N$).
   Most of the screening comes from free carriers in graphene, and it becomes less efficient as $N$ increases and the distance between graphene and GaSe increases.
   The increase in the Fr\"ohlich EPI coming from polar-optical phonons in BN is more drastic because the bare Fr\"ohlich potentials from each BN layer add up.}
   \label{fig:num_layers}
\end{figure}

Fig.\ \ref{fig:num_layers} shows the effect of  the number $N$ of BN layers.
As $N$ increases, so does the intensity of the remote Fr\"ohlich EPIs, as contributions from each layer add up.
The efficiency of remote screening from graphene is limited to momenta smaller than a certain critical $q$, which decreases with increasing distance $d$ between GaSe and graphene.
This effect can be roughly estimated as follows.
Both the bare Fr\"ohlich potentials felt by graphene and the induced potential from graphene felt by GaSe decay as $e^{-qd}$.
Remote screening becomes inefficient when $e^{-q \times (2d)}\ll 1$, that is when $q \gg \frac{1}{2d}$, where $d = (N+1) \times  3.4$ \AA\ in our model.
Thus, from an electrostatics standpoint, it is better to minimize the number of layers.
Of course, what is feasible and optimal in a practical device may depend on other parameters.\\

\begin{figure}[h]
   \includegraphics[width=0.45\textwidth]{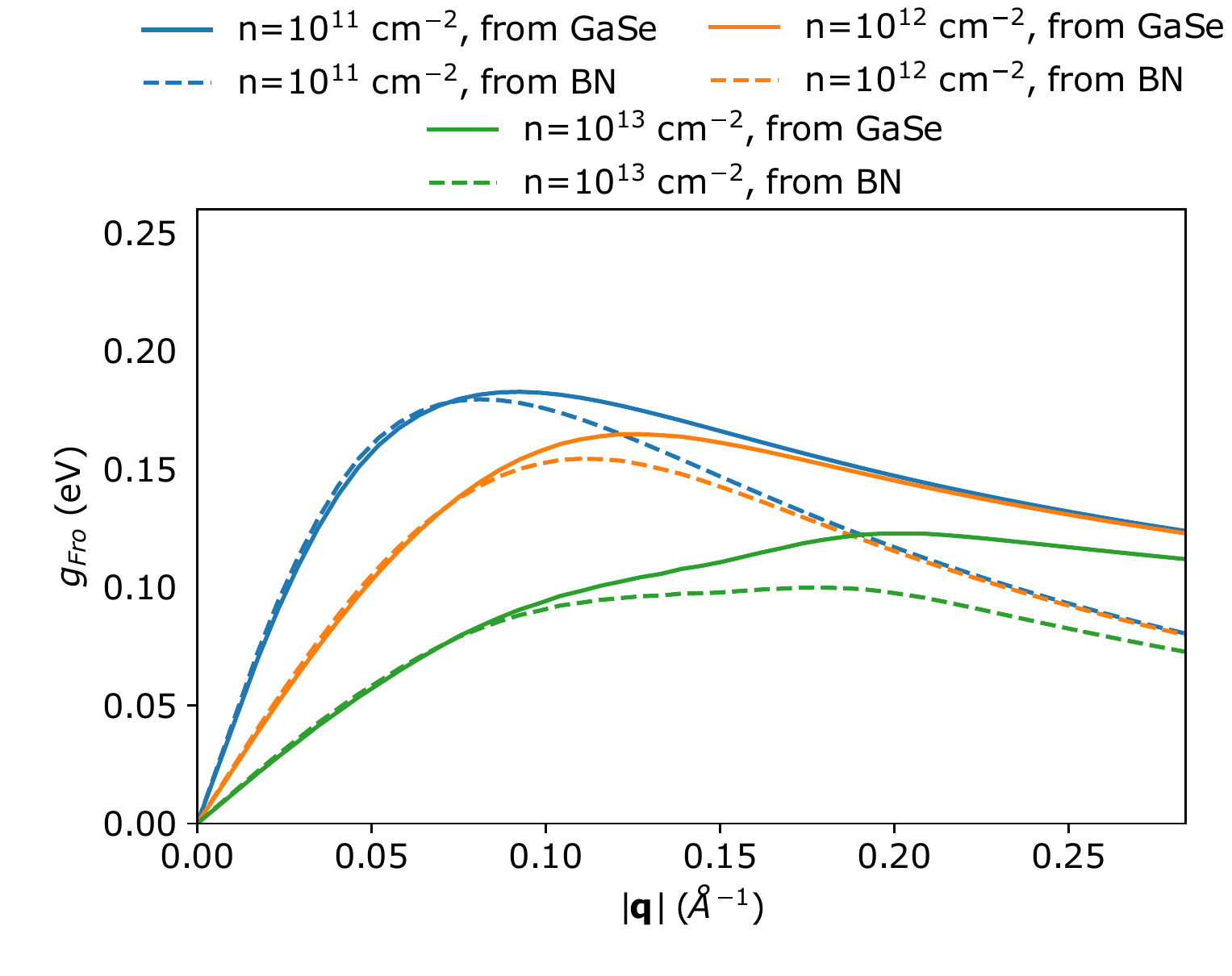}
   \caption{Fr\"ohlich EPI from polar-optical phonons in both GaSe (plain) and BN (dashed), as felt by electrons in GaSe, in GaSe($n$)/BN/Gr(doped), changing the doping in GaSe $n$.
   As the carrier density increases, free-carrier screening from electrons in GaSe comes into play, ensuring the ability of the system as a whole to screen EPIs at the larger momenta involved in transport.}
   \label{fig:doping}
\end{figure}

Fig.\ \ref{fig:doping} shows the variation of the EPIs in GaSe($n$)/BN/Gr(doped) with respect to carrier density $n$ in GaSe.
Screening from the free carriers added in GaSe is modeled as described in App.\ \ref{app:free} at room temperature.
First, note that as $n$ increases, the Fermi surface gets larger, and the momenta most relevant to transport ($q \approx 2 k_F$) increase.
The efficiency of intrinsic free-carrier screening also follows the size of the Fermi surface and extends to larger $q$. As a result, the Fr\"ohlich EPIs (intrinsic and remote) at momenta relevant for transport are always significantly screened.

\begin{figure}[h]
   \includegraphics[width=0.45\textwidth]{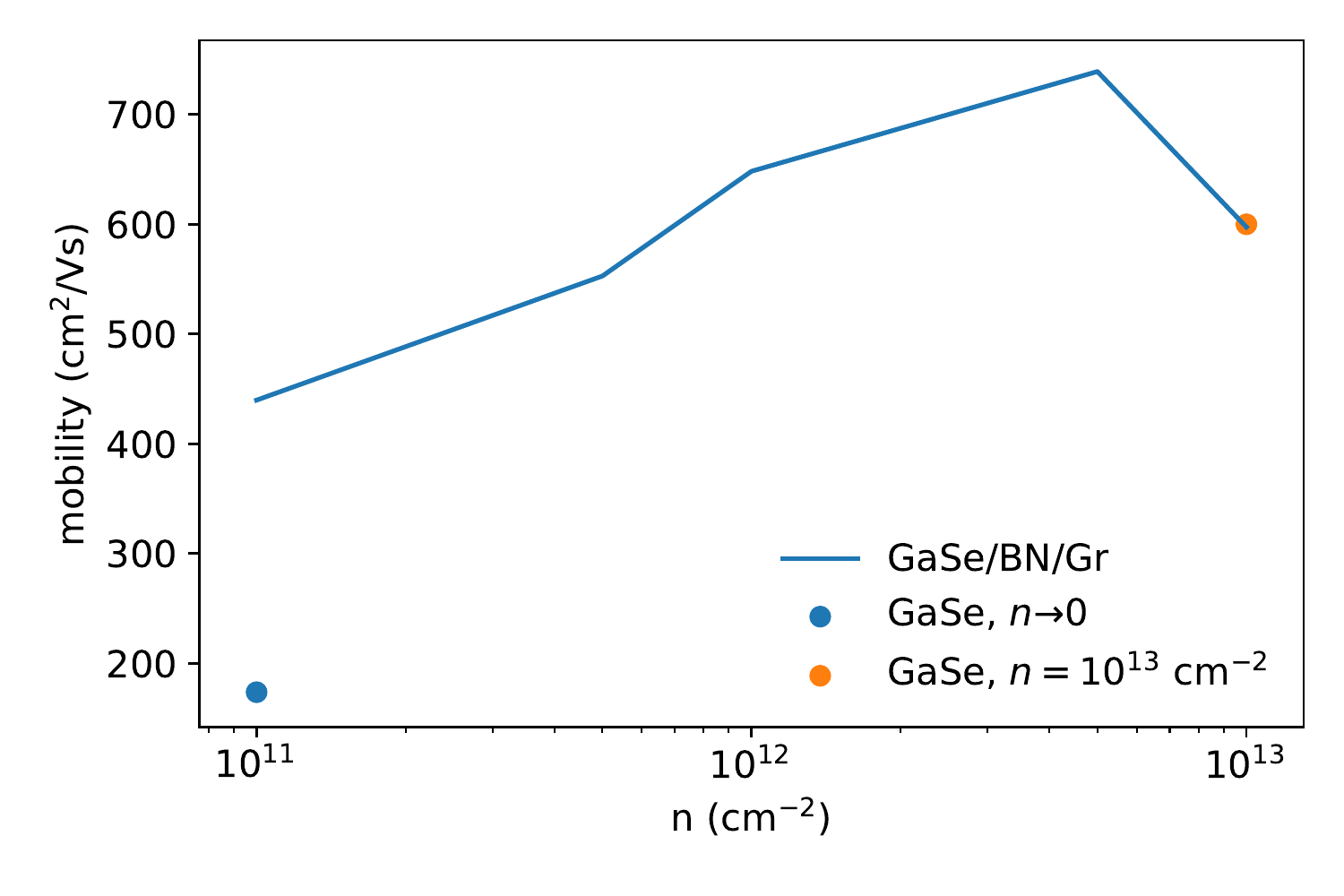}
   \caption{Mobility versus electrostatic doping in the GaSe($n$)/BN/Gr(doped) system, replacing $g_{LO}$ by our momentum and doping dependent model for the Fr\"ohlich EPIs within the VdW heterostructure. Remote EPIs from BN are also added. The dots are direct, full DFPT calculations for a monolayer GaSe in the low doping limit (EPIs computed in neutral GaSe, transport solved at $n=10^{11}$ cm$^{-2}$) and at high doping ($n=10^{13}$ cm$^{-2}$).}
   \label{fig:mobility}
\end{figure}

In Fig.\ \ref{fig:mobility}, we inject the modelled Fr\"ohlich EPIs of the last system, GaSe($n$)/BN/Gr(doped), back into the Boltzmann transport solver and look at the mobility as a function of GaSe doping $n$.
The rest of the EPIs, with a much smaller contribution to transport (see Fig.~\ref{fig:mob_neut}), are taken from DFPT calculation in doped ($n = 10^{13}$ cm$^{-2}$) single layer GaSe.
Those other EPIs, and in particular those from acoustic phonons, are only partially screenable.
This complicates their modelling in the current framework, which is left for future work.
Here, for acoustic EPIs, we essentially replace the doping-dependent screening  from the whole VdWh by the screening from GaSe at $n = 10^{13}$ cm$^{-2}$.
Note that the mobility predicted by this model at $n = 10^{13}$ cm$^{-2}$ is very close to the one computed directly in DFPT in GaSe alone, at the same doping.
After determining that the contribution of remote Fr\"ohlich EPIs from BN is negligible (as expected due to the high frequency of the associated phonons), we deduce that this agreement is the result of the cancellation of two effects: additional screening from graphene which tends to increase the mobility, and the overestimation of the bare Fr\"ohlich coupling discussed in appendix \ref{app:Frolich} which tends to reduce it.
Both contributions were estimated to be on the order of $10\%$ of the mobility (with opposite signs). Since the overestimated bare Fr\"ohlich coupling is just a limitation of our model, we expect the current results to be conservative.
In addition to preserving good performance at high doping, the benefits of the VdWh are obvious, since instead of degrading towards the neutral isolated limit at low doping ($<200$ cm$^{-2}$/Vs), the mobility stays relatively high above $450$ cm$^{-2}$/Vs.
As carrier density decreases, the intrinsic free-carrier screening lost from depleting carriers in GaSe is compensated by remote free-carrier screening from graphene.
Remote screening thus extends the outstanding performance of GaSe to the full range of doping typically achievable experimentally.

\section{Conclusion}
We have developed a semi-analytical model to simulate the electrostatic response of any VdWh.
The model is parametrized via the DFPT density-response of each individual layer to a monopole and dipole perturbation potential.
We use this model to explore the possibility of using metallic 2D layers in VdWh (e.g. doped graphene) to induce free carrier screening remotely in a current carrying semiconducting layer.
This is particularly relevant for 2D semiconductors with transport properties limited by screenable EPIs such as the Fr\"ohlich interaction.
In particular, such materials would typically showcase excellent transport performance in presence of free-carrier screening (e.g.\ at high-doping), but lower mobilities in its absence (low doping).
Using GaSe as a prototypical example, we show that integrating it in a VdWh device with doped graphene as a remote screener and BN as a separator enhances the mobility at low doping.
The mobility is thus maintained at a  consistently high value of $\sim 500-600$ cm$^{-2}$/Vs on a wide range of carrier concentrations.

\begin{acknowledgements}
We are grateful to Matteo Calandra for his help with the initial implementation of the linear response to a monopole perturbation in the Quantum ESPRESSO code.
We acknowledge that the results of this research have been achieved in DECI project RemEPI on ARCHER EPCC with support from the PRACE aisbl.
Simulation time was also awarded by PRACE (project id. 2020225411) on MareNostrum at Barcelona Supercomputing
Center - Centro Nacional de Supercomputación (The Spanish National Supercomputing Center)".
Computational resources have been provided by the Consortium des Equipements de Calcul Intensif (CECI), funded by the Fonds de la Recherche Scientifique de Belgique (F.R.S.-FNRS) under Grant No. 2.5020.11 and by the Walloon Region.
T.S.\ acknowledges support from the University of Liege under Special Funds for Research, IPD-STEMA Programme.
M.G.\ acknowledges support from the Italian Ministry for University and Research through the Levi-Montalcini program and from the Swiss National Science Foundation through the Ambizione program.
M.J.V.\ gratefully acknowledges funding from the Belgian Fonds National de la Recherche Scientifique (FNRS) under PDR grant T.0103.19-ALPS.
\end{acknowledgements}

\appendix

\section{Model details}
\label{app:model}

This appendix details technical aspects of the electrostatics model: the extraction and parametrization of each monolayer's response from DFPT; the semi-numerical scheme to combine those responses and solve the electrostatics of the VdWh; the model for the perturbing Fr\"ohlich potential; the inclusion of doping-induced free-carrier screening.
A quick comparison of this model with the existing QEH method \cite{Andersen2015,Gjerding2020} is also provided.

\subsection{Monolayer response from DFPT}
\label{app:resp}

\begin{figure*}[h]
   \includegraphics[width=0.95\textwidth]{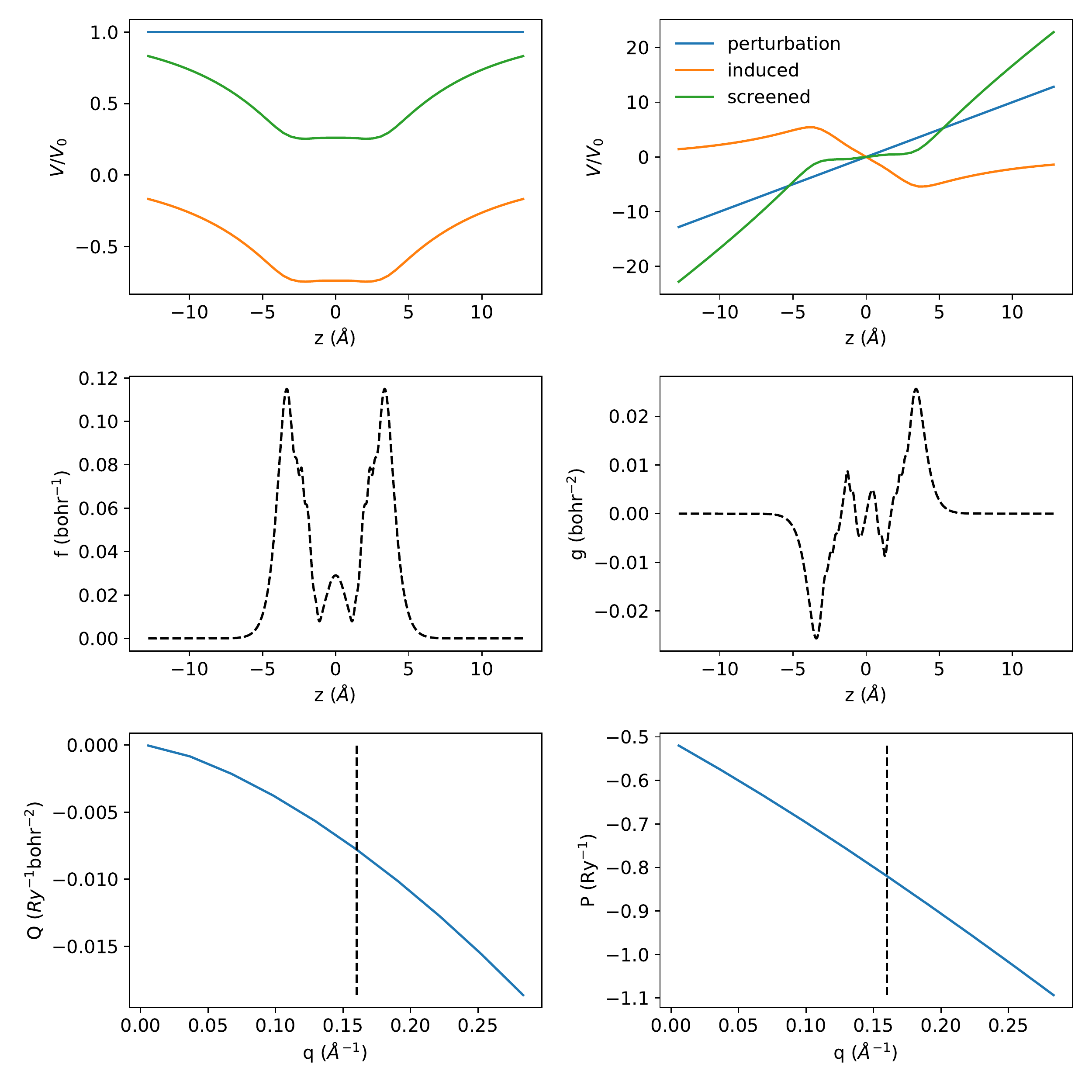}
   \caption{The response of single-layer GaSe is extracted from DFPT.
   Left and right panels concern the monopole and dipole contributions to the response, respectively. The top panels show the bare perturbation (fixed at unity value), as well as the induced potential and the screened potential (sum of perturbation and induced potential).
   The middle panels show the normalized out-of-plane profile functions $f$ and $g$.
   The bottom panels show the momentum-dependent amplitudes of the responses $Q$ and $P$.
   The dashed line corresponds to the momentum at which the potentials and profile functions are plotted.}
   \label{fig:model}
\end{figure*}

In-plane periodicity and symmetry suggests we Fourier transform quantities in-plane and keep the out-of-plane real-space variable: $(x,y,z) \to (\mathbf{q},z)$. Since the systems are assumed isotropic in the plane, we further simplify and use $q = |\mathbf{q}|$.
The response of the layer to a generic perturbing potential is written as:

\begin{equation}\label{eq:Vind}
\begin{aligned}
V_{\rm{ind}}(q,z) &= v_c(q) \int e^{-q |z-z'|} \delta n(q, z') dz' \\
\delta n(q, z) &= \int \chi(q,z,z') V_{\rm{ext}}(q,z') dz'
\end{aligned}
\end{equation}

where $v_c(q)=\frac{2\pi e^2}{q}$ is the Coulomb kernel in 2D and the (interacting) response function $\chi$ is written as in Eq.~\eqref{eq:chi}.
The profile are normalized as follows:
\begin{align}\label{eq:fg}
\int f(q,z-z_0) dz = \int (z-z_k)g(q,z-z_k) dz =1
\end{align}
for a layer $k$ centered around $z_k$.
The response from each layer is computed within DFPT.
The layers are perturbed by a constant potential to probe the monopole response, then by a linear one (constant field) to probe the dipole response:
\begin{align}\label{eq:Vpert}
V_{\rm{ext}} (q,z) = V_{\rm{p, mono}}(q,z) &= V_0 \\
V_{\rm{ext}} (q,z) = V_{\rm{p, dip}}(q,z) &= V_0 z
\end{align}
where $V_0$ is a small quantity. The density response $\delta n(q,z)$ is first extracted in reciprocal space $\delta n(q, G_z)$, with as many $G_z$ as the energy cutoff and supercell size dictates, then Fourier transformed back in real space, onto $1000$ points $z \in [-c/2,c/2]$.
Injecting the above perturbing potential into Eqs.~\eqref{eq:Vind} with $\chi$ from \eqref{eq:chi},  the density response to the monopole perturbation (renormalized by $V_0$) gives $Q(q) f(q,z-z_0)$ while the dipole one gives $P(q) g(q,z-z_0)$.
Those quantities are computed on $10$ $\mathbf{q}$-points in the range of values relevant for transport ($0 - 0.26$ \AA$^{-1}$), and interpolated (scipy, quadratic spline) on finer sets of $q$-points.
The $Q,P$ parameters are defined as the integrals over $z$ and $f,g$ as the normalized profiles.
Fig.\ \ref{fig:model} shows those quantities in GaSe.

\subsection{VdWh electrostatics}
\label{app:combi}

Within the random phase approximation (RPA)~\cite{RPA1,RPA2,RPA3}, the linear density response to an external potential can be approximated as the non-interacting response to the screened (external plus induced) potential
\begin{equation}\label{eq:deltan-chi0}
\delta n (q,z) = \int \chi_0 (q,z,z')\left[ V_{\rm ext} (q,z') + V_{\rm ind} (q,z')\right]dz'
\end{equation}
where $\chi_0(q,z,z')$ is the non-interacting response function. When considering a VdWh, if the overlap between the wavefunctions in different layers can be neglected, we can write the full non-interacting response as the sum of contributions from each layer $\chi_0(q,z,z')=\sum_k \chi_0^{(k)}(q,z,z')$ and thus also the density response can be decomposed as the sum of layer contributions
\begin{equation}
\delta n (q,z) = \sum_k \delta n^{(k)} (q,z)
\end{equation}
with $\delta n^{(k)} (q,z)$ given by Eq.~\eqref{eq:deltan-chi0} with $\chi_0$ replaced with $ \chi^{(k)}_0$. Within the RPA, this will be the main approximation to build the VdWh response, which underlies also other approaches~\cite{Andersen2015,Gjerding2020,Guo2020}.

In particular, within this approximation also the potential induced by the heterostructure is the sum of the potentials induced by each layer $k$:
\begin{align}
V_{\rm ind}(q,z) = \sum_k v^k_{\rm ind}(q,z)
\end{align}
Each layer  thus responds to an effective external potential that is the sum of the external potential and the potentials induced in each of the other layers. We can thus write
\begin{align}
\delta n^{k} (q,z) &= \int \chi_0^k (q,z,z')\left[ V_{\rm ext} (q,z') +\sum_{m} v^m_{\rm ind} (q,z')\right]dz' \\
&= \int \chi^k (q,z,z')\left[ V_{\rm ext} (q,z') +\sum_{m\ne k} v^m_{\rm ind} (q,z')\right]dz' \notag
\end{align}
where $\chi^k$ is the interacting response function of the $k$-th layer. By using Eq.~\eqref{eq:Vind}, we then have
\begin{align} \label{eq:vindk}
v^k_{\rm ind}(q,z) =& v_c(q) \int e^{-q |z-z'|}  \int \chi^{k}(q,z',z'')
\bigg[ V_{\rm{ext}}(z'') \\
&+ \sum_{m \ne k} v^m_{\rm ind}(q,z'') \bigg]
dz'' dz' \nonumber
\end{align}
We replace $\chi^k$ by its expression Eq.~\eqref{eq:chi}, and re-write the result as a system of $2N$ equations with $2N$ unknowns, where $N$ is the number of layers in the heterostructure and the unknowns are the induced potentials averaged on each layer using either $f$ or $g$ as weight.
\begin{widetext}

Injecting $\chi^k$ from Eq.~\eqref{eq:chi} into Eq.~\eqref{eq:vindk}, we obtain:

\begin{align} \label{eq:vindk2}
v^k_{\rm ind}(q,z) = v_c(q)  \left(
Q_k(q) F_k(q, z-z_k) \left[ \bar{v}^{ext}_{k}(q)+\bar{v}_{k}(q) \right]
+ P_k(q) G_k(q, z-z_k) \left[ \bar{w}^{ext}_{k}(q)+\bar{w}_{k}(q) \right]
\right)
\end{align}

with
\begin{align}
  F_k(q,z) &=  \int e^{-q |z-z'|} f_k(q,z') dz' \\
  G_k(q,z) &=  \int e^{-q |z-z'|} g_k(q,z') dz'
\end{align}
$\bar{v}, \bar{w}$ designate projections/averages of the potentials over the monopole or dipole profiles:
\begin{align}
  \bar{v}^{ext}_k(q) &=  \int  f_k(q,z') V_{\rm{ext}}(z') dz' \\
  \bar{w}^{ext}_k(q) &=  \int g_k(q,z') V_{\rm{ext}}(z') dz' \\
  \bar{v}_k(q) &=  \int  f_k(q,z') \sum_{m \ne k} v^m_{\rm ind}(q,z') dz' \\
  \bar{w}_k(q) &=  \int g_k(q,z') \sum_{m \ne k} v^m_{\rm ind}(q,z') dz'
\end{align}

Finally, multiplying both sides of Eq.~\eqref{eq:vindk2} by $f_p(q,z-z_p)$ or $g_p(q,z-z_p)$, integrating over $z$, and summing over $k \ne p$, one obtains a set of equations that $\bar{v}_p(q)$ and $\bar{w}_p(q)$ should satisfy at each $q$:

\begin{align}
\bar{v}_p(q) &= v_c(q) \sum_{k \ne p} \left(
Q_k(q) C[f_p,f_k](q) \left[ \bar{v}^{ext}_{k}(q)+\bar{v}_{k}(q) \right]
+ P_k(q) C[f_p,g_k](q) \left[ \bar{w}^{ext}_{k}(q)+\bar{w}_{k}(q) \right]
\right) \\
\bar{w}_p(q) &= v_c(q) \sum_{k \ne p} \left(
Q_k(q) C[f_p,f_k](q) \left[ \bar{v}^{ext}_{k}(q)+\bar{v}_{k}(q) \right]
+ P_k(q) C[g_p,g_k](q) \left[ \bar{w}^{ext}_{k}(q)+\bar{w}_{k}(q) \right]
\right)
\end{align}

where $C[f_p,g_k](q)$ denotes the following double integral:
\begin{align}
  C[f_p,g_k](q) &=  \int \int f_p(q,z-z_p) e^{-q |z-z'|} g_k(q,z'-z_k) dz'dz \\
\end{align}
and similarly for other combinations of $f,g$ functions with indices $p,k$.

Thus, we obtain a set of $2N$ equations with $2N$ unknowns, where $N$ is the number of layers.
This is solved numerically at each momentum to find $v_k(q),w_k(q)$ for each layer $k$.
The full $z$-dependency of the induced potentials are then recovered via Eq.~\eqref{eq:vindk2}.

Within this fornalism, one can easily compute layer-specific static dielectric functions. The system is perturbed by the monopole potential of Eq.~\eqref{eq:Vpert}, and the dielectric function on layer $k$ is defined as:
\begin{align}
  \epsilon_k(q) & = \frac{\int V_{\rm p,mono}(q,z) f_k(q,z-z_k) dz}{\int (V_{\rm p,mono}(q,z)+V_{\rm ind}(q,z)) f_k(q,z-z_k) dz} = \frac{1}{\int (1+V_{\rm ind}(q,z)/V_0) f_k(q,z-z_k) dz}
\end{align}
Similarly, the dielectric function of the entire VdWh is defined as:
\begin{align} \label{eq:epsdef}
  \epsilon(q) & = \frac{1}{\int (1+V_{\rm ind}(q,z)/V_0) \frac{1}{N}\sum_k f_k(q,z-z_k) dz} = \frac{1}{ \frac{1}{N}\sum_k \frac{1}{\epsilon_k(q)}}
\end{align}

\end{widetext}

\subsection{Fr\"ohlich perturbation}
\label{app:Frolich}

Since the screening is computed within the VdWh via the electrostatics model, only the bare Fr\"ohlich potential generated by polar-optical phonons is necessary here.
In previous works \cite{Sohier2016}, we assumed a generic square profile for the polarization density that is the source of the Fr\"ohlich potential.
Here we assume that the polarization follows the profile characterizing the materials' dielectric response found in DFPT $f$.
The \textit{bare} Fr\"ohlich potential from layer $k$ centered around $z_k$ is then:
\begin{align} \label{eq:Frobare}
V_{\rm{Fro}, k} (q,z) &= C^k_Z A \int e^{-q|z-z'|} f(q,z'-z_k) dz'
\end{align}
where $A$ is the area of the unit cell.
For a layer centered around $z_k$,  $C^k_Z$ is defined in each layer as:
\begin{align}
C^k_Z  = \frac{2 \pi e^2}{A} \sum_a \frac{ \mathbf{e}_{\mathbf{q}} \cdot Z_a^k \cdot \mathbf{e}^a_{\Gamma, LO}}{\sqrt{2M_a \omega_{\Gamma, LO}}}
\end{align}
where $a$ is an atomic index, $\mathbf{e}_{\mathbf{q}} = \mathbf{q}/|\mathbf{q}|$, $Z_a^k$ are the Born effective charges of atom $a$ in layer $k$, $M_a$ is the mass of atom $a$, and $\omega_{\Gamma, LO}, \mathbf{e}^a_{\Gamma, LO}$ are the frequency and eigenvector of the LO mode at $\Gamma$.
Note that we include an extra electron charge factor $e$ in the definition of the potential.
Strictly speaking, $V$ is the electric potential \textit{energy} of a test charge $e$ within the potential.
The variation of $C_Z$ as a function of $\mathbf{q}$ due to the phonon eigenvector and frequency are neglected : the values at $\Gamma$ are used.
DFPT calculations in the GaSe and BN give $C_Z = 1.124$ and $1.994$ eV, respectively.
The bare Fr\"ohlich potential is then a sum of the potentials from each ``activated'' layer, usually one material at a time, $V_{\rm{Fro}} (q,z) = \sum_k V_{\rm{Fro}, k} (q,z)$.
To compute the corresponding electron-phonon coupling strength on layer $m$, the screened potential (bare Fr\"ohlich plus induced) is then averaged over the profile of layer $m$:

\begin{align} \label{eq:Froav}
g_{\rm{Fro}, m}(q) = \frac{1}{A}\int (V_{\rm{Fro}} (q,z) + V_{\rm{ind}} (q,z)) f_m(q,z-z_m) dz.
\end{align}

\begin{figure}[h]
   \includegraphics[width=0.45\textwidth]{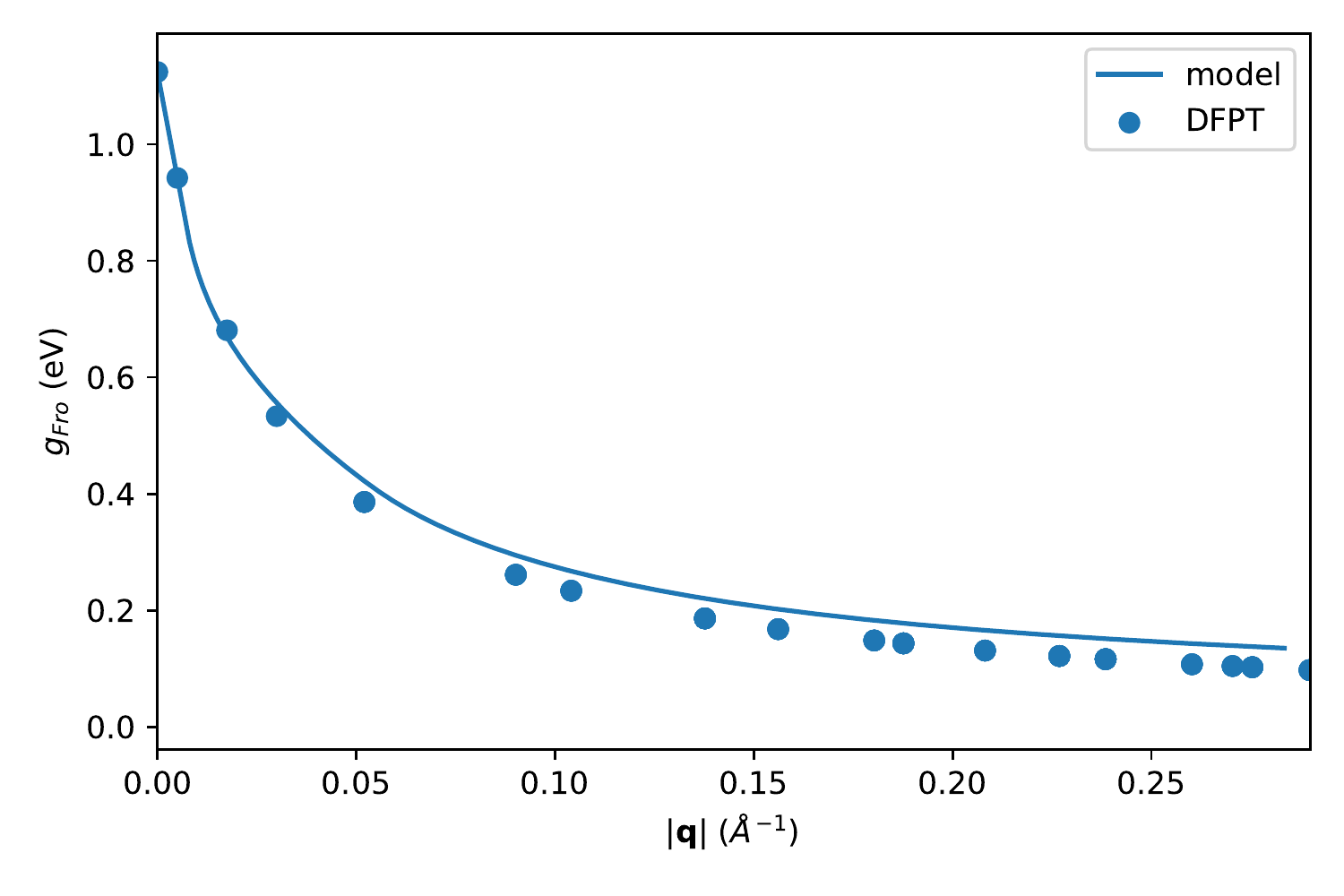}
   \caption{Comparison of DFPT and modelled Fr\"ohlich electron-phonon coupling in neutral, single-layer GaSe, as a function of phonon momentum. In this system the bare Fr\"ohlich interaction is screened by the dielectric response of GaSe. Since this response is reproduced virtually exactly by our model (see appendix \ref{app:tests}), the discrepancy seen here comes from the model for the bare Fr\"ohlich potential (Eq.~\eqref{eq:Frobare}) and the use of the profile function $f$ rather than the wavefunctions to obtain the electron-phonon coupling matrix elements (Eq.~\eqref{eq:Froav}).}
   \label{fig:Fro_model_limits}
\end{figure}

Beyond the long wavelength $q \to 0$ limit, this model is an approximation. More precisely, the form of the $q$-dependency in Eq.~\eqref{eq:Frobare} is an assumption, and Eq.~\eqref{eq:Froav} is a simplification. To obtain the electron-phonon coupling matrix elements, one should indeed project the potential perturbation on the wavefunctions of the initial and final electronic states. While relatively good, this simple model does not reproduce the DFPT coupling exactly. As shown in Fig.~\ref{fig:Fro_model_limits}, the model fits very well at small $q$, but it overestimate the coupling by $\sim 20\%$ at $q = 0.15$~\AA, which is the size of the largest, high-doping Fermi surface considered here.

\subsection{Free-carrier screening}
\label{app:free}

The following is necessary to model free-carrier screening from the carrier density injected into the operating material, GaSe.
The non-interacting density response function\cite{Ma2014a,Maldague1978,Gjerding2020} is evaluated numerically using the band structure of the materials on a fine grid of $\mathbf{k}$-points ($96 \times 96$ for GaSe):
\begin{align}
\chi_0(T, q) &= -  \frac{2 FF(q)}{(2\pi)^2}  \int d^2\mathbf{k} \frac{n_{FD}(\varepsilon_{\mathbf{k}})-n_{FD}(\varepsilon_{\mathbf{k}+\mathbf{q}})}{\varepsilon_{\mathbf{k}+\mathbf{q}}-\varepsilon_{\mathbf{k}}}
\end{align}
where $n_{FD}$ is the Fermi-Dirac occupation (which also depends on the chemical potential and the temperature) and the form factor FF(q) is computed using the monopole profile function $f$.
\begin{align}
FF(q) = \int f(q,z-z_0) \int e^{-q|z-z'|} f(q,z'-z_0) dz dz'
\end{align}
Note that this form factor relates the $Q(q)$ parameters to an (interacting) response function $\tilde{\chi}(q)$ that would be integrated in the out-of-plane direction and would include only the monopole response $\tilde{\chi}(q) =  Q(q) FF(q)$.
The free-carrier response is combined with the dielectric response already in the model within RPA.
In practice, the non-interacting response functions are added $\tilde{\chi}_0 = \tilde{\chi}^d_0+\tilde{\chi}^f_0$ ($d$ for dielectric, $f$ for free-carrier), the interacting response function is recomputed as $\tilde{\chi}(q) = \tilde{\chi}_0(q)/(1 - v_c(q) \tilde{\chi}_0(q))$ and a new $Q(q) = \tilde{\chi}(q)/FF(q)$ is injected in the solver.

\subsection{Tests}
\label{app:tests}

We compare direct DFPT and the model in single and bilayer BN, for the range of momenta considered in this work (relevant for transport). Similar tests were done on single layer GaSe and graphene. Of course, direct DFPT calculations on the full heterostructure are too complex and expensive, which is the premise of this work. Nevertheless, we expect that bilayer BN captures the model's main approximation in the context of this work, that is neglecting the interlayer hybridization.

\begin{figure*}[h]
    \includegraphics[width=0.95\textwidth]{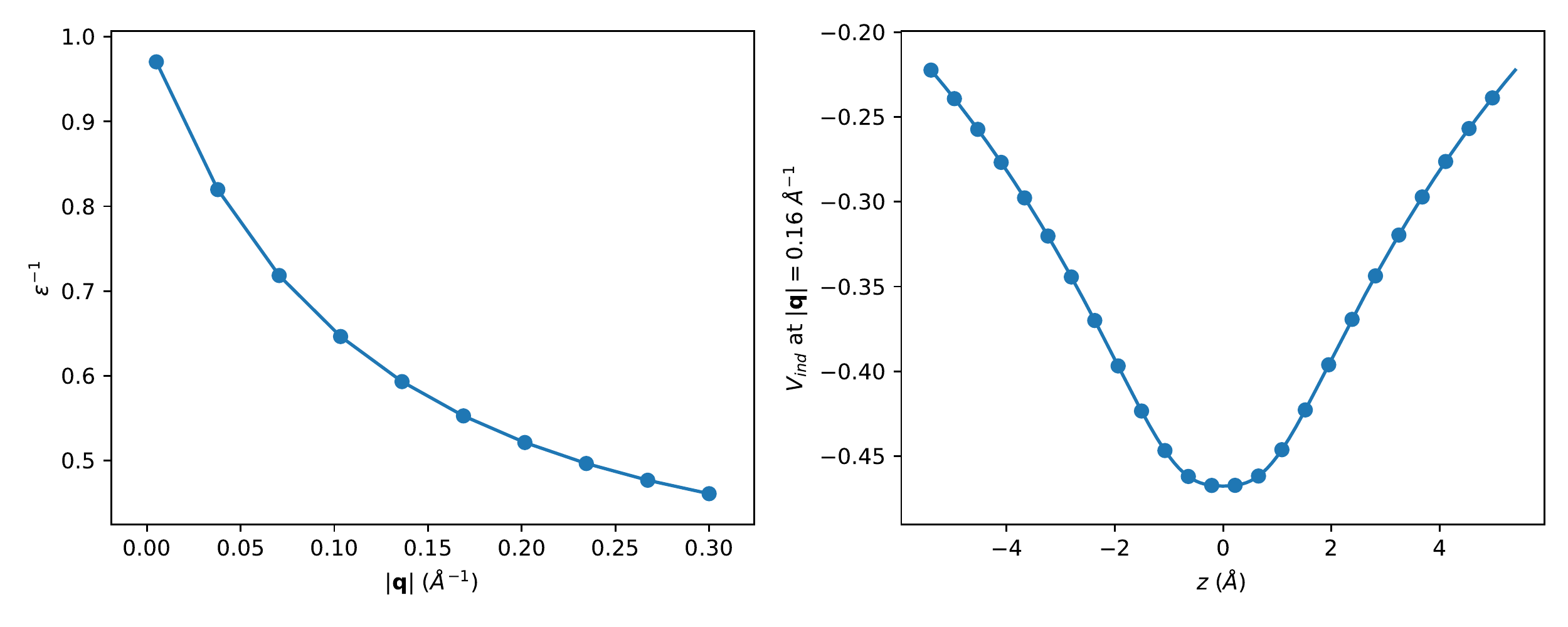}
    \includegraphics[width=0.95\textwidth]{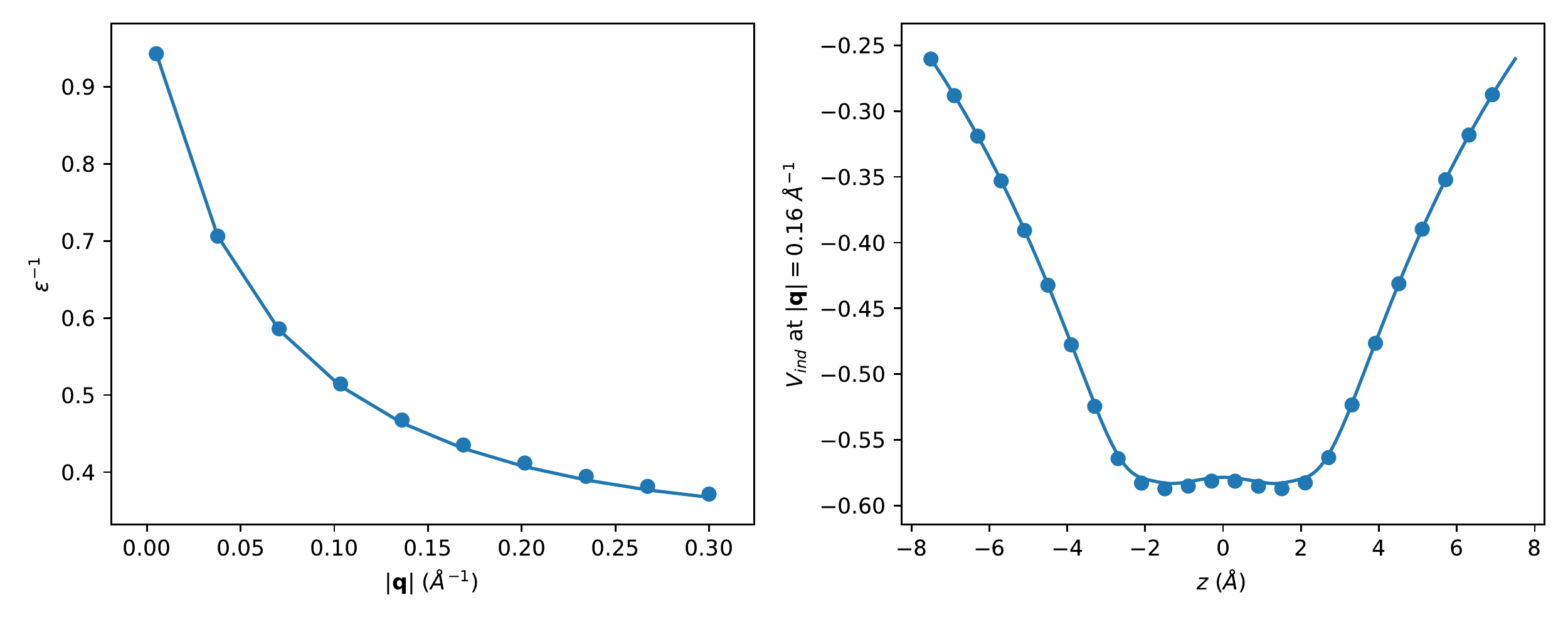}
    \includegraphics[width=0.95\textwidth]{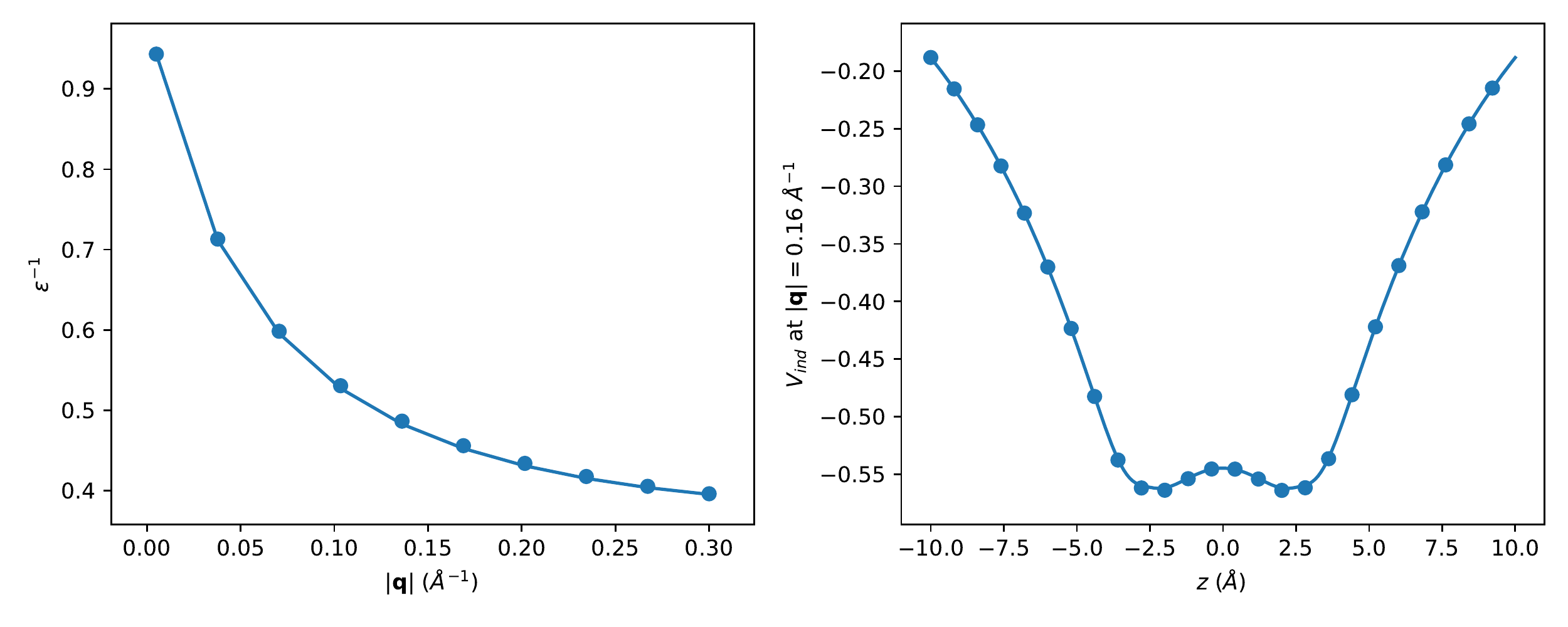}
    \caption{Comparison of DFPT and our model for the electrostatic response of single layer BN (top), bilayer BN with a 3.4 A interlayer distance (middle), bilayer BN with a 5 A interlayer distance (bottom). Left is the inverse dielectric response function, as a function of the perturbation momentum, as defined in Eq. \eqref{eq:epsdef}. Right is the induced potential, as a function of the out-of-plane space variable, for a given momentum of the perturbation.
}
    \label{fig:tests}
 \end{figure*}

In single layer BN, top of Fig.~\ref{fig:tests}, the agreement is perfect, which is expected since our model is built to reproduce exactly the response of a single layer.
In bilayer BN with $3.4$~\AA\  interlayer distance, middle panels of Fig.~\ref{fig:tests}, the agreement is excellent, but some small discrepancy can be detected. We see the effects of our model's approximations.
To confirm that the main effect is associated with the interlayer hybridization, we look at  bilayer BN with $5$~\AA\ interlayer distance, bottom panels of Fig.~\ref{fig:tests}.
Indeed, as the interlayer distance is increased, the validity of the approximation increases, and the (already small) discrepancy is further reduced.

\subsection{Comparison with long wavelength analytical models}
\label{app:AnaCompa}

\begin{figure*}[h]
   \includegraphics[width=0.9\textwidth]{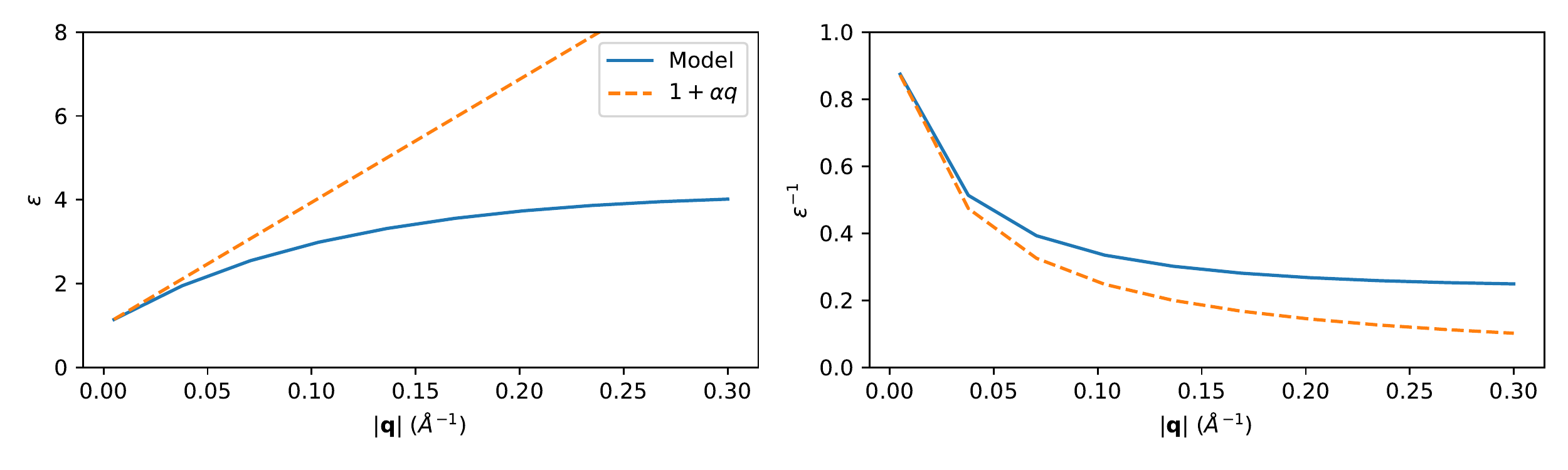}
   \includegraphics[width=0.9\textwidth]{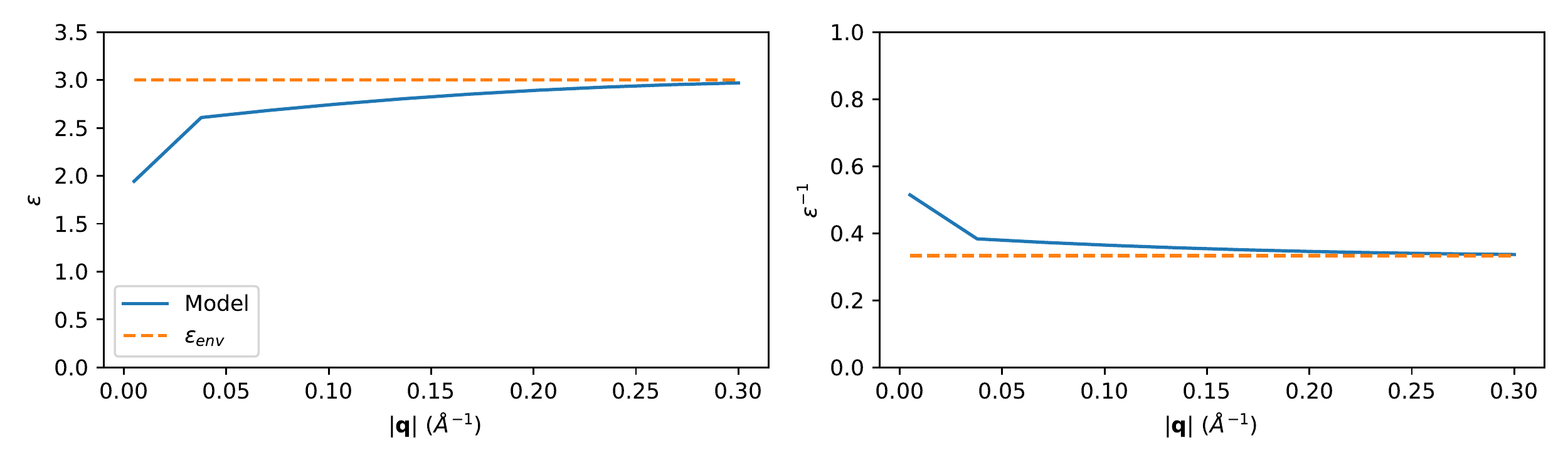}
   \includegraphics[width=0.9\textwidth]{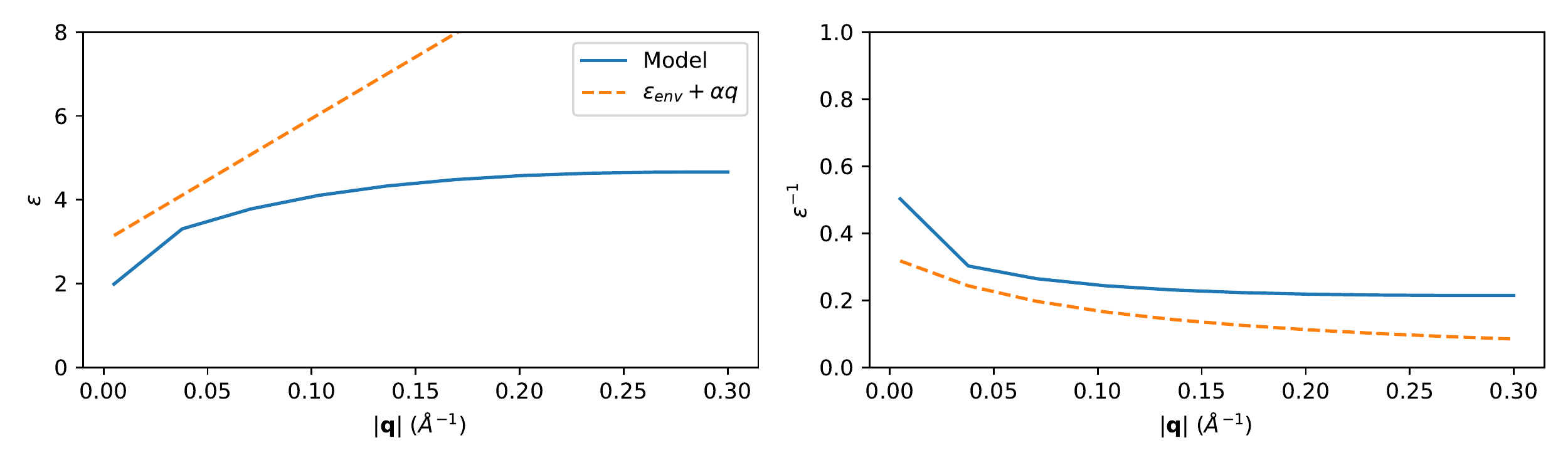}
   \caption{The dielectric function and its inverse, as computed in the VdWh electrostatics model, are compared to simpler, long-wavelength analytical models. Top: single layer GaSe. Middle: GaSe on top of 40 layers of BN, dielectric function of the outermost BN layer. Bottom: GaSe on top of 40 layers of BN, dielectric function of the GaSe layer.}
   \label{fig:alpha1}
\end{figure*}

Here we compare dielectric functions computed within our model (see Eq.~\eqref{eq:epsdef}) with simple analytical models that are expected to describe the long wavelength behavior of screening. Unless noted otherwise, the dielectic function we consider is the layer-specific one relating to GaSe.
At the level of the monolayer, the long wavelength linearized model of the dielectric function
$\epsilon = 1 + \alpha q$, where $\alpha$ is the polarizability of the 2D layer~\cite{Cudazzo2011,Latini2015,Trolle2017,Tian2020}, is routinely employed~\cite{Keldysh1978,Cudazzo2011,Berkelbach2013,Sohier2017a,Chen2020}.
It captures only the first order term (in $q$) of the dielectric function, while our model goes well beyond, as shown Fig.~\ref{fig:alpha1} (top) for GaSe.
We now consider substrate effects.
They are relatively easily included in the long wavelength analytical model as follows: $\epsilon = \epsilon_{env} + \alpha q$, with  $\epsilon_{env} = (\epsilon_s + 1)/2$ being the average dielectric constant of the environment made of substrate, $\epsilon_s$, on one side and vacuum on the other.
In the VdWh electrostatics model, we use 40 layers of BN to simulate the substrate.
The above analytical expression is for a semi-infinite substrate.
The 40 layers system should behave similarly when $q > 2 \pi/t ~ 0.05$~\AA$^{-1}$ ($t$ being the thickness of our “substrate”).
We first check that assumption by plotting in Fig.~\ref{fig:alpha1} (middle) the epsilon as felt by the outer BN layer (furthest from GaSe).
If the slab was infinitely thick, one should obtain $\epsilon_{env}$. In reality, this starts to be indeed a reasonable approximation for $q>0.05$~\AA$^{-1}$.
Then, in Fig.~\ref{fig:alpha1} (bottom), we plot the dielectric function on the GaSe layer, and compare with the simple model  $\epsilon = \epsilon_{env} + \alpha q$, with alpha being around $30$~\AA (rather large) in GaSe.
Here the simple analytic model fails at all wavevectors:
i) at small $q$ because of the limited thickness of our BN substrate. That would be attenuated if the number of layers is increased. However, in the context of VdWhs, 40 layers is probably a reasonable number.
ii) at large $q$ because the linearized long-wavelength version of GaSe's dielectric function fails. Note that ``large'' $q$ here (say from $0.05$ to $0.15$~\AA$^{-1}$) is still relevant for transport.

Thus, in the current framework (vdWh, few layers, and the range of momenta relevant for transport plotted above), simple long wavelength analytical models are clearly not enough, similar to what happens also in the context of exciton binding energies~\cite{Latini2015,Trolle2017}.
There are more complex and accurate expressions in the literature~\cite{Trolle2017,Sohier2016}, which would probably work much better in the semi-infinite substrate case.

\begin{figure*}[h]
  \includegraphics[width=0.9\textwidth]{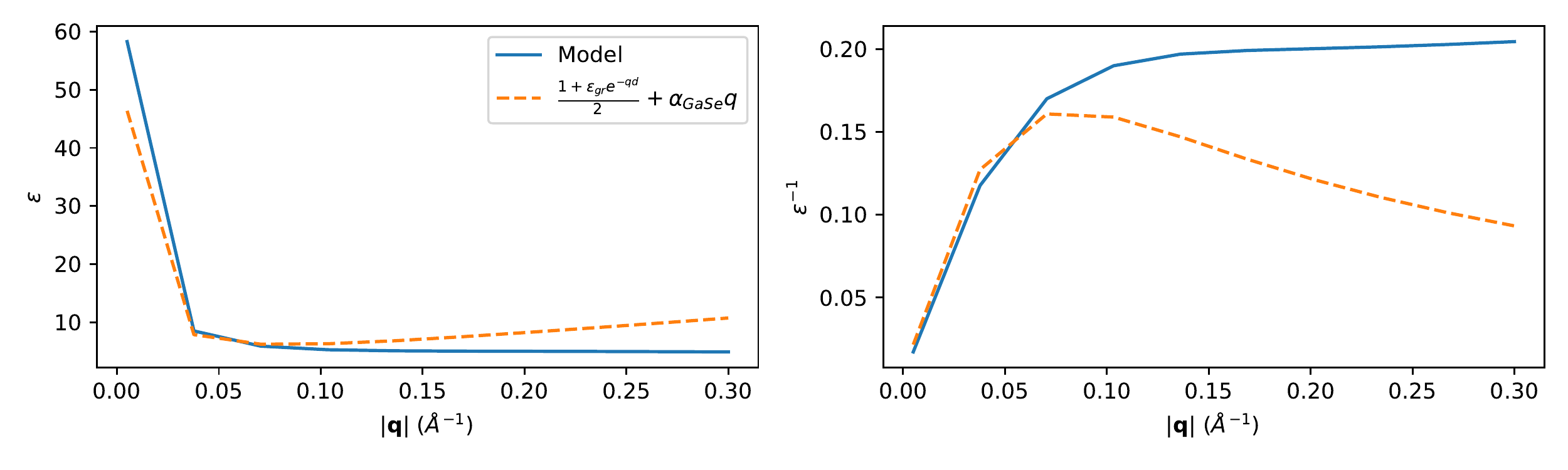}
   \includegraphics[width=0.9\textwidth]{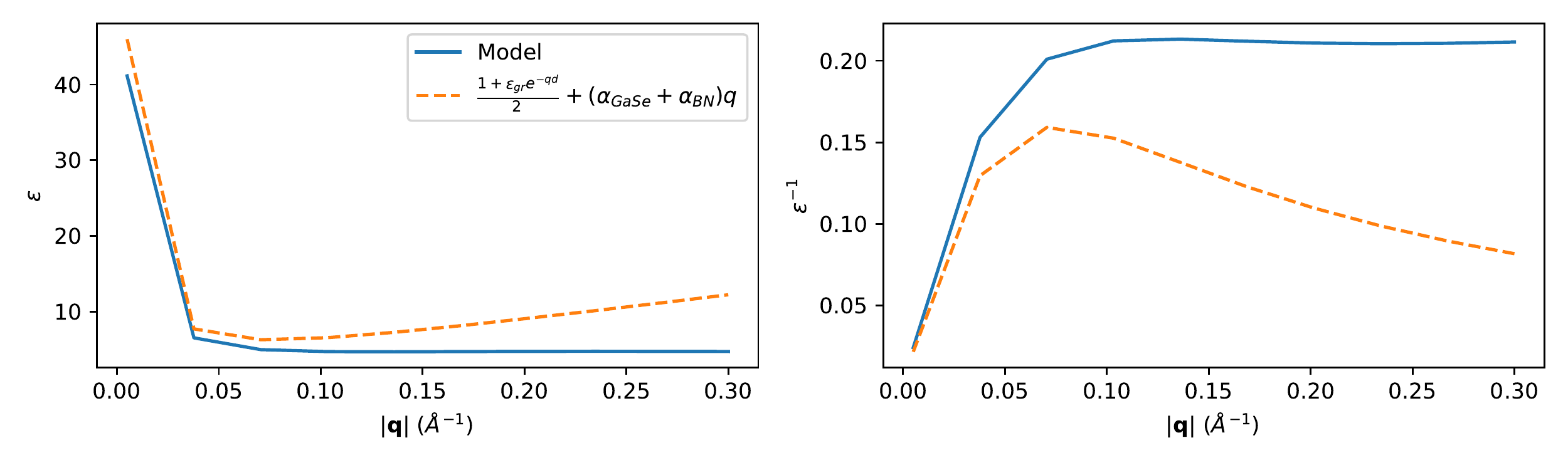}
   \caption{The dielectric function and its inverse, as computed in the VdWh electrostatics model, are compared to simpler, long-wavelength analytical models. Top: GaSe and graphene, dielectric function of GaSe. Bottom: GaSe, BN and graphene, dielectric function of GaSe.}
   \label{fig:alpha2}
\end{figure*}

Finally, we investigate whether at least the remote metallic screening can be easily captured with a simple long-wavelength model.
We first consider a simpler VdWh, made of GaSe and graphene. A simple, relatively naive expression would be $\epsilon =  (1+ \epsilon_{\rm gr} e^{-qd})/2 + \alpha q $, where $d$ is the distance between the two layers.
The agreement in Fig.~\ref{fig:alpha2} (top) is decent at small wavevectors, when the response is dominated by metallic graphene. The large $q$ regime is still problematic for the same reasons as before.

In Fig.~\ref{fig:alpha2} (bottom), we consider a more realistic system with BN as a separator, and compare to the analytical formula: $\epsilon =  (1+ \epsilon_{\rm gr} e^{-qd})/2+ (\alpha_{\rm GaSe} + \alpha_{\rm BN}) q $.
The agreement is still decent in the $q\to0$ limit, while larger  wavevectors are still poorly captured.
A more accurate analytical model (beyond first order in $q$) might be possible for systems such as GaSe/BN/Gr, mixing semiconducting and metallic responses, although the analytical derivations start to become tedious.
The VdW heterostructures could also be much more complicated, and then a semi-analytical model becomes absolutely necessary.

\begin{figure*}[h]
   \includegraphics[width=0.45\textwidth]{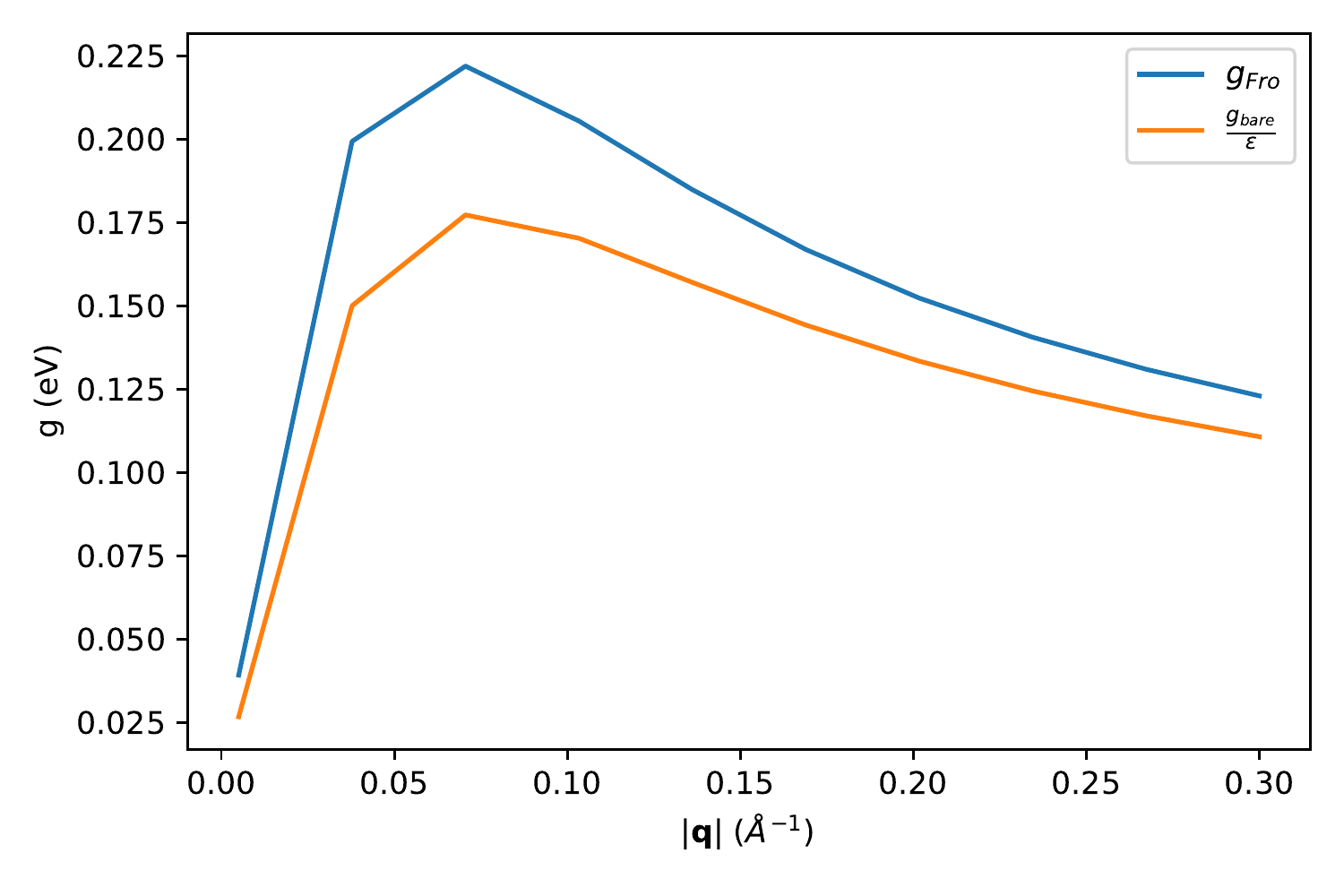}
   \caption{The screened Fr\"ohlich electron-phonon interaction, computed either by simulating the response of the GaSe/BN/graphene to the bare Fr\"ohlich potential, or by screening the bare Fr\"ohlich potential with the dielectric function.}
   \label{fig:alpha3}
\end{figure*}

Finally, note that in this work we are interested mostly in the screening of a Fr\"ohlich potential.
The screened Fr\"ohlich potential could naively be obtained by dividing the bare Fr\"ohlich potential by $\epsilon$.
However, all the above dielectric functions are computed in response to a monopole perturbation, Eq.~\eqref{eq:Vpert}.
The Fr\"ohlich perturbation is different: it is centered on a given layer and decays as $e^{-q|z|}$ in the out-of-plane direction.
This brings added complexity: each layer is effectively perturbed by a Frohlich potential of different magnitude, and has a dipolar component that is absent from all the analytical models considered above, but captured in our semi-analytical formulation.
Even if $\epsilon$ is exact, these aspects will produce important differences, as shown below. We first compute $\epsilon_{\rm GaSe}$ in the standard way, by perturbing the GaSe/BN/Gr system with a monopole potential (constant in the out-of-plane direction).
Then we perturb with a Fr\"ohlich potential and compare $g_{\rm Fro}$ as defined in Eq.~\eqref{eq:Froav}
with $g_{\rm bare}/\epsilon$ where $g_{\rm bare} = \int V_{\rm{Fro}} (q,z)  f_m(q,z-z_m) dz$.
For a single isolated layer the two quantities coincide because the perturbation as seen by the layer generating the potential is almost constant.
However, Fig.~\ref{fig:alpha3} shows that for a VdWh the discrepancy is manifest, bringing additional errors if one were to assume $g_{\rm Fro} \approx g_{\rm bare}/\epsilon$ in an analytical model.

\subsection{Comparison with QEH model}
\label{app:QEHcompa}

The QEH model~\cite{Andersen2015,Gjerding2020} is another approach to combine the independent responses of each layer to compute the response of a VdWh. It relies on the same approximations as the present case, i.e.\ it is an RPA approach where the density response is written as the sum of contributions of each layer. Although the QEH model and the present strategy share several similarities, we summarize here the main differences:
i) In QEH, the linear density response function is computed as in Ref.~\onlinecite{Yan2011} then projected on two potential profile functions for monopole and dipole contributions.
In our approach, the full density response to monopole and dipole perturbing potentials is directly interpolated on the $\mathbf{q}$-points in which we are interested. By avoiding the projection on basis functions for the potentials, we keep the full $z$-dependency of the induced potentials during the process of solving the electrostatics. In QEH, a 1D Poisson equation in the out-of-plane direction needs to be solved at the end of the process to recover this $z$-dependency.
ii) In QEH, the combination of the responses is achieved within the Dyson equation formalism. We instead map the problem onto a linear system of equations, with a  somewhat simpler physical interpretation. As far as we can tell, the formalisms are equivalent on this point.
iii) QEH allows for a dynamical treatment of the responses, while our method is presently limited to the static limit.
iv) QEH enables the treatment of anisotropic materials, while we presently stay in the isotropic case.
v) Concerning free carrier screening, we arrived independently at a very similar approach (free carrier screening was added only very recently in the QEH method\cite{Gjerding2020}).
Still, there are some practical differences in the calculations. QEH makes use of the quadratic band approximation to compute $\chi_0$, while it is evaluated numerically on the full band structure in our case. Also, we account for form factor effects. We do not expect a significant difference to arise from those aspects, at least for light doping and simple band structures.

Overall,  although the two methods rely on the same approximations and both target the combined response of an heterostructure, they were developed with different quantities and applications in mind, which explains the differences above.
Specifically, our approach is centered around potentials and their variations as a function of momentum $q$ and out-of-plane position $z$.
The system can easily be perturbed by an arbitrary potential (Fr\"ohlich in this work), and the response of each individual layer as well as the total effective potential can be easily extracted, bringing a clear picture of the underlying electrostatics.
QEH focuses on more macroscopically averaged quantities like the dielectric function and the underlying potentials are less accessible.
In particular, its application to Fr\"ohlich perturbing potentials, although technically possible, might suffer from limitations associated with the projection on linear or constant profiles (see point i above), while in the present approach an average over a physically motivated material- and $q$-dependent profile is adopted.
We have also checked for quantities that are instead easily accessible in both methods, like the static dielectric function, our framework agrees with the QEH, taking multilayer BN as a prototypical example.

\clearpage

\bibliographystyle{myapsrev}
\bibliography{bib}

\end{document}